\documentclass[a4paper,fleqn,usenatbib]{mnras}
\usepackage[T1]{fontenc}
\usepackage{ae,aecompl}
\usepackage{amsmath}	
\usepackage{amssymb}	
\usepackage{epsfig}

{\newif\ifnotend
\notendtrue
\def\veclist{ABCDEFGHIJKLMNOPQRSTUVWXYZabcdefghijklmnopqrstuvwxyz.}
\def\top#1#2.{#1}
\def\tail#1#2.{#2.}
\loop\expandafter\xdef\csname v\expandafter\top\veclist\endcsname%
{{\noexpand\bf\expandafter\top\veclist}}
\edef\veclist{\expandafter\tail\veclist}
\if\veclist.\notendfalse\fi\ifnotend\repeat}

\newcommand {\sigpar}{{\sigma_{\rm p}}}
\newcommand {\sigparg}{{\sigma_{\rm p,g}}}
\newcommand {\vpar}{{\rm p}}
\newcommand {\delpar}{{\delta_{\rm p}}}
\newcommand {\vlos}{v_{\rm los}}
\newcommand {\kms}{\,{\rm km}\,{\rm s}^{-1}}
\newcommand {\magn}{\,{\rm mag}}
\newcommand {\mas}{\,{\rm mas}}

\newcommand {\brp}{{\rm G_{BP}}}
\newcommand {\grp}{{\rm G}}
\newcommand {\rrp}{{\rm G_{RP}}}

\newcommand {\pc}{\,{\rm pc}}
\newcommand {\kpc}{\,{\rm kpc}}

\newcommand {\gl}{{l}}
\newcommand {\gb}{{b}}
\newcommand {\eclat}{\beta}
\newcommand {\mul}{\mu_l}
\newcommand {\mub}{\mu_b}

\newcommand {\becl}{\beta}
\newcommand {\nvis}{{\rm n_{\rm vis}}}
\newcommand {\Ug}{U_g}
\newcommand {\Vg}{V_g}
\newcommand {\Wg}{W_g}
\newcommand {\Vc}{V_c}
\newcommand {\Uh}{U}
\newcommand {\Vh}{V}
\newcommand {\Wh}{W}

\newcommand {\siglos}{\sigma_{\rm los}}

\newcommand {\degs}{\,{\rm deg}}
\newcommand {\B}[1]{{\boldsymbol{#1}}}
\newcommand {\Rsun}{{R_{0}}}

\newcommand {\vsun}{{\B{\upsilon_\odot}}}
\newcommand {\Usun}{{U_{\!\odot}}}

\newcommand {\Wsun}{{W_{\!\odot}}}
\newcommand {\Vsun}{{V_{\!\odot}}}

\newcommand {\Tvw}{T_{\rm vw}}

\newcommand {\Tuw}{T_{\rm uw}}
\newcommand {\Tww}{T_{\rm ww}}

\newcommand {\zsun}{\,{\rm z}_\odot}

\title[Distances and parallax bias in Gaia DR2]{Distances and parallax bias in Gaia DR2}
\author[R. Sch\"onrich et al.]{Ralph Sch\"onrich$^1$\thanks{E-mail: ralph.schoenrich@physics.ox.ac.uk},
        Paul McMillan$^2$, and Laurent Eyer$^3$\\
        $^1$University of Oxford, Rudolf Peierls Centre for Theoretical Physics, Clarendon Laboratories, Oxford, OX1 3PU, UK\\
        $^2$Lund Observatory, University of Lund, S{\"o}lvegatan 27, 221 00 Lund, Sweden\\
        $^3$Department of Astronomy, University of Geneva, Chemin des Maillettes 51, 1290 Versoix, Switzerland         
        }

\date{Draft, \today}
\pubyear{2019}

\begin{document}
\label{firstpage}
\pagerange{\pageref{firstpage}--\pageref{lastpage}}
\maketitle

\begin{abstract}
We derive Bayesian distances for all stars in the RV sample of Gaia DR2, and use the statistical method of Sch{\"o}nrich, Binney \& Asplund (2012) to validate the distances and test the Gaia parallaxes. In contrast to other methods, which rely on special sources, our method directly tests the distances to all stars in our sample. We find clear evidence for a near-linear trend of distance bias $f$ with distance $s$, proving a parallax offset $\delpar$. On average, we find $\delpar = -0.054 \mas$ (parallaxes in Gaia DR2 need to be increased), when accounting for the parallax uncertainty under-estimate in the Gaia set (compared to $\delpar = -0.048 \mas$ on the raw parallax errors) with negligible formal error and a systematic uncertainty of about $0.006 \mas$. The value is in concordance with results from asteroseismic measurements, but differs from the much lower bias found on quasar samples. We further use our method to compile a comprehensive set of quality cuts in colour, apparent magnitude, and astrometric parameters. Last, we find that for this sample $\delpar$ appears to strongly depend on $\sigpar$ (when including the additional $0.043 \mas$) with a statistical confidence far in excess of $10 \sigma$ and a proportionality factor close to $1$, though the dependence varies somewhat with $\sigpar$. Correcting for the $\sigpar$ dependence also resolves otherwise unexplained correlations of the offset with the number of observation periods $\nvis$ and ecliptic latitude. Every study using Gaia DR2 parallaxes/distances should investigate the sensitivity of their results on the parallax biases described here and - for fainter samples - in the DR2 astrometry paper.
\end{abstract}


\section{Introduction}
This paper presents the first statistical evaluation of parallax biases for a general stellar sample and the first derivation of unbiased stellar distances for the subsample of Gaia DR2 with line-of-sight velocity measurements \citep[][]{GaiaDR2, Katz19}. Other than most other approaches, our method tests the distances of all stars in the selected sample and not just specific subgroups with different physical properties. 

Since the first release of Gaia DR2 data, there has been clear evidence for offsets in the parallax measurements. In the data release papers, \cite{Lindegren18} found a general offset of Gaia parallaxes by $\delpar = -0.029 \mas$ (in the sense that the quoted values were too low) when studying known quasars in the Gaia catalogue. In addition, the median offset of the quasars is patchy on the sky \citep[see also Fig. 15 in][]{Arenou18}, and thus we also have to expect an error term $\sigpar$, which will in many aspects behave like a random error (as long as these patches are not resolved). The calibration involved in the data analysis for the astrometric solution of Gaia depends on both effective wavelength (colour) and magnitude (directly and through the 'window class'), so both offsets/biases and random errors cannot be expected to be homogeneous along the entire magnitude and colour range of the sample. The quasar catalogue, which is the best direct benchmark for Gaia parallaxes, is very faint (typically at apparent magnitudes $\grp > 17$), and the colour distribution does not match the stellar colour distribution very well. In contrast their internal checks between different parts of the analysis show significant changes in behaviour in particular at magnitudes $\grp \sim 16, 13, 10$ \citep[caused by their analysis windows, see Fig. 16 in][]{Lindegren18} and changes in their astrometric measurement accuracy (likely near $\grp \sim 8$, see their Fig. 9). In addition, sources in several remote objects, in particular the LMC display a finely structured pattern in sky position of mean offsets related to the scanning law of Gaia.

In \cite{Helmi18} the same offset can be seen when comparing the globular cluster mean parallaxes to the more accurate (at small parallax) values from the \cite{Harris96} catalogue. \cite{Helmi18} also show that the Large and Small Magellanic clouds, and most of the satellite dwarf spheroidal galaxies they study, have negative average parallaxes. Consistent with these general offsets, but differing in the magnitude of the effect, \cite{Stassun18} found an offset of $(-0.082 \pm 0.033) \mas$ for a sample of $89$ very bright ($G < 12 \magn$) binary systems, however with a large scatter of the single measurements. Yet, given the patchiness of the offset, and the very different magnitude range probed, these values could well be in agreement. Just when we were about to submit this paper, the latest study of \cite{Graczyk19} found a value of $\delpar = -(0.054 \pm 0.024) \mas$, which by an amusing coincidence is exactly our result, albeit with a much larger uncertainty. We also note that binarity was not worked into the astrometric solution, so it is to be expected that binary systems carry a different bias from the main stellar sample. Similarly, comparison with asteroseismic values points to significant parallax underestimates \citep[][]{Zinn18}, though again for rather specific subsets of stars in magnitude and colour/stellar evolutionary stage. However, their result of $\delpar = -0.05 \mas$ bears high confidence, and again differs from the quasar result. \cite{Sahlholdt18} looked at a smaller sample of dwarfs with asteroseismic data, and the offset they found was closer to that of the quasar sample, but with more significant uncertainty ($-0.035\pm0.016\mas$). No test so far has measured what we are really concerned about: the bias of the full stellar sample and how it depends on the other properties of the observation.

Unbiased distances are key to virtually every problem in modern astrophysics, and given the large sample sizes, we now need these distances at the $1\%$ level. For example the wave-pattern discovered by \cite{SD18}, which was confirmed by \cite{Huang18} and \cite{Kawata18}, and which is likely related to the later findings of \cite{Antoja18}, has a total amplitude of below $1 \kms$. To compare this: even at a perfect location towards the Galactic anticentre, the solar reflex motion will translate into a $\sim 0.1 \kms$ bias for every $1 \%$ in mean distance bias. Larger and more complex bias patterns will arise at other sky positions from cross-correlations between the velocity component measurements.

This work offers a solution to this problem. We will derive unbiased distances to the RV subsample (about $7$ million stars at magnitudes $\grp \lesssim 15$) using the method proposed in \cite{SA17}. By measuring the selection function directly from the sample, we derive a data-informed, and thus nearly unbiased prior, which suits a sample better than model-based priors \citep[e.g.][]{Astraatmadja16, BailerJones18} designed to fit pre-existing models of the entire Gaia sample, e.g. \cite{BailerJones18}. As discussed in \cite{SA17}, mismatch or neglect of the selection function will result in systematic bias of similar size to the measurement uncertainties for individual stars (i.e. of order $20\%$ for the common parallax quality cut of $\vpar / \sigma_p > 5$). Our previous results have established that this method gives an unbiased translation from parallaxes to distances. Consequently, with these distances we can now directly measure and correct biases in the parallaxes using the statistical method of \cite{SBA}. 

Our paper is structured as follows: we start with a short description of the used data and coordinate system definitions in Section \ref{sec:Datadef}, followed by a description of our statistical distance estimator in Section \ref{sec:Theory}. After this, we provide the formalism for deriving Bayesian distances in Section \ref{sec:Distances}, including a derivation of the distance-dependent selection function $S(s)$. In Section \ref{sec:Measurement} we quantify the different parallax biases in Gaia DR2, followed by a comparison to previous distance derivations and a comment on the distance to the Pleiades. Section \ref{sec:qualcuts} provides a summary of quality cuts necessary in Gaia DR2, followed by the Conclusions.

\section{Data and definitions}\label{sec:Datadef}

\subsection{Coordinate frame and definitions}\label{sec:definitions}

Throughout this paper, we will use the standard definitions for Galactic coordinates and the Local Standard of Rest. We employ Galactic cylindrical coordinates $(R,z,\phi)$, where $R$ is the in-plane distance to the Galactic centre, $z$ is the altitude above or below the Galactic midplane, and $\phi$ is the Galactic azimuth, with the Sun placed at $\phi = 0$. The distance of a star to the Galactic centre is termed $r = \sqrt{R^2 + z^2}$, for the solar galactocentric distance, we use the value $\Rsun = 8.27 \kpc$ from \cite{S12}, which is also in agreement with other determinations \citep{Gillessen09,McMillan17}, and only slightly in tension with the latest estimates from measurements of stellar orbits around Sgr A* from \cite{Gravity18}. The vertical displacement of the Sun from the midplane, $\zsun = 0.02 \kpc$ is taken from \cite{Joshi07}. We also tested that it does not have any significant impact on our results. The velocity vector in the heliocentric,\footnote{We use the somewhat negligent Galactic dynamicists' term "heliocentric", while in truth, Gaia is measuring quantities in the solar system baryocentric frame. With relative motions between the two frames below $0.1 \kms$, this difference does not matter.} Cartesian frame is defined as $(\Uh,\Vh,\Wh)$ with a right-handed set of components pointing radially inwards, in the direction of Galactic rotation, and upwards perpendicular to the plane. The velocity vector components in the Galactocentric cylindrical frame are analogously termed $(\Ug,\Vg,\Wg)$. To translate these velocity components, we use the motion of the Sun against the Local Standard of Rest as measured in \cite{S10,S12}: $(\Usun,\vsun,\Wsun) = (11.1,250,7.24) \kms$, and if necessary, use the azimuthal velocity of the Sun against the Local Standard of Rest ($\Vsun = \vsun - \Vc = 12.24\kms$). For simplicity's sake, we call $p$ the parallax of a star and $\sigpar$ the effective uncertainty of the parallax measurement assumed in that instance, which, depending on the examined set of assumptions, may contain the additional $\delta \sigpar = 0.043 \mas$ added in quadrature to the Gaia pipeline value $\sigparg$. 

\subsection{Data}\label{sec:data}

Here we use the Gaia RV sample \citep[][]{Cropper18, Katz18} from Gaia DR2 \citep[][]{GaiaDR2} with more than $7$ million stars that have both astrometric and line-of-sight velocity measurements from the onboard spectrograph \citep[][]{Sartoretti18} on the Gaia spacecraft \citep[][]{Prusti16}. To ensure the quality of the data, we apply, if not stated otherwise for a specific task, a few quality cuts that were discussed in data release papers, e.g. \cite{Lindegren18}, namely: the number of visibility periods $\nvis \ge 5$ to ensure a full astrometric solution, a parallax quality cut of $\vpar/\sigma_p > 5$ and lower limit $\vpar > 0.1 \mas$, which translates to approximately demanding a distance $s \lesssim 10 \kpc$, an excess noise smaller than $1$, line-of-sight velocity limits of $|\vlos| < 550 \kms$ and $\siglos < 10 \kms$. We usually remove the Galactic mid-plane from our sample, i.e. require $|\gb| > 10 \degs$. For our statistics the Galactic mid-plane carries no signal and we thus avoid problems with crowding and excessive reddening. We checked, though, that the measurement of the distance prior from low-$|b|$ data is similar to our higher latitude main sample, and provide distance estimates for these stars in the derived catalogue. In previous papers \citep[see][]{SA17} we uncovered major problems with $\vlos$ measurements, in particular with LAMOST. Here, we just note that our tests of $\vlos$ accuracy and precision looked very decent on the Gaia sample, and we will concentrate on the more pressing issue of distances and Gaia parallaxes. We follow the convention of Gaia papers to call their apparent magnitudes in the three broad colour bands $(\brp, \grp, \rrp)$. Problems with capitalisation conventions do not arise, since we do not discuss absolute magnitudes through most of the paper.

\section{Statistical Distance Estimation}\label{sec:Theory}

\subsection{General thought}\label{sec:Generalthought}

For the determination of distance bias, we rely on the method of \cite{SBA} (hereafter SBA), which has been applied on various samples. The method relies on correlations between velocities, which depend on the position on the sky. The estimator is readily derived by writing down an estimate of stellar kinematics while allowing for a systematic distance bias $f = \langle s'/s \rangle$, where $s'$ denotes the estimated distance, $s$ the real distance to a star. This $f$ affects both tangential velocity components simultaneously, correlating them. To explain this with a simple example: imagine approaching a mountain horizontally. Knowing your velocity, your mind automatically has a clear estimate of the distance to the summit. This is because any incorrect estimate would translate your horizontal motion into a vertical component of motion of the mountain, i.e. the top of the mountain would have to be growing or shrinking (if you had over- or under-estimated the distance, respectively). Analogously, all parts of the mountain base below your level would appear to be moving downwards (upwards). And your brain knows this is not usually what mountains do. The SBA method allows us to extend this intuition and make it more robust against assumptions (i.e. we do not assume any mean motion or fixed velocity ellipsoid) to stellar samples.

The strength of the SBA method now lies in directly using the spatial dependence of this correlation of the heliocentric velocity components on  galactic longitude ($\gl$) and latitude ($\gb$). As long as we have a sufficient sky coverage, we do not depend on classic assumptions of other methods: think of our mountain example. Observing many mountains around us, we gain a significant advantage in control of systematics over the use of just one single mountain. In a simplified picture, we just measure the pattern of apparent vertical velocities of all the mountain tops and mountain bases around us and try to find the distance correction that makes the angular dependence of this pattern disappear. Misjudging our own velocity, we would equally bias the motion of all summits and mountain roots, leaving the statistical distance evaluation unscathed. Similarly, we do not care about our horizontal velocity, since it can i) be measured and ii) would just affect our prediction for the strength of the effect around us; however, we just seek the distance factor at which the correlation of vertical velocity components with sky position disappears, making our own horizontal motion irrelevant. Translated to our real problem: Assumptions about the solar velocity do not matter for our method.

Similarly, let us assume that we are sitting in the middle of an orogeny (or reading Calvino's {\it Cosmicomics}) and both mountain summits and roots are rapidly rising and sinking into the ground. Observing just one mountain in front of us, we would indeed infer that distances are overestimated, but behind us, the correlation term reverses sign, i.e. the apparent distance underestimate there cancels out the distance over-estimate inferred from the opposite direction. We learn from this that typically modes of the disc cancel out in a sample with large sky-area. Analogously, a wide halo stream passing through would cancel by the spatial terms. In short, galaxy physics can only affect our statistical measurements, if they vary across the sky in a way that correlates with the angle-terms of our method. In most cases (e.g. global breathing modes, streams) they will cancel out at first order. 

Last, we note that other than our imaginary mountains, stars move horizontally, so in addition to the spatial correlations, we can benefit from two different horizontal velocity components with different dependence on sky position.

\subsection{Formal argument and specific implementation}

The formal derivation of our method (see SBA for a stringent treatment) is done by simply writing down what happens in the measurement. The vector of observed values $(s'\mul,s'\mub,\vlos)$, where $s'$ is the observationally inferred distance, $\mul, \mub$ are the proper motions in Galactic longitude $\gl$ and latitude $\gb$.  This vector is translated into the measured velocity components $(U,V,W)$ by a matrix $\vM$ depending on $\gl$ and $\gb$. Since this is an orthogonal matrix, the inverse mapping (i.e. from the original velocity components) is done with the transverse $\vM^t$. If we now assume that distances are changed by some relative bias 
\begin{equation} 
f = (s' - s)/s $,$
\end{equation}
where $s$ is the real distance, we can relate:
\begin{equation}
\left( \begin{matrix}
\Uh \\  \Vh \\ \Wh
\end{matrix} \right) = \vM(\vI + f\vP)\vM^t 
\left( \begin{matrix} 
\Uh_0 \\ \Vh_0 \\ \Wh_0 
\end{matrix} \right)
\end{equation}
where the index $0$ indicates the real values, and $P$ is $\rm{diag}(1,1,0)$, which projects to the two proper motion components. Now, we see that the observed velocity components are correlated by $f$ via the matrix $\vT = \vM \vP \vM^t$. Since the equations are linear, the average $f$ can thus be gained by a similar linear regression of any target velocity component $v_i$ onto the other velocity components $v_j$ multiplied with $T_{ij}$, the components of the matrix $\vT$.

As discussed in SBA, using the in-plane velocity components $(\Uh, \Vh)$ mixes the statistics with a Galactic rotation estimate, and given the very large sample size here, we make the choice to avoid this possible source of systematic bias. We thus limit this study to using the correlation of both $U$ and $V$ velocity components with the vertical motion $W$. The relevant part of $T$ is thus
\begin{equation}
\left( \begin{matrix}
\Tuw \\ \Tvw \\ \Tww
\end{matrix} \right) = \left( \begin{matrix}
\cos{\gl}\sin{\gb}\cos{\gb} \\ \sin{\gl}\sin{\gb}\cos{\gb} \\ 1 - \cos{\gb}^2
\end{matrix} \right)
\end{equation}
Our method applies corrections for the following biases of this measurement:
\begin{itemize}
\item $\vlos$ determination errors, $\siglos$, which would appear as distance underestimates (typically negligible due to the excellent precision and accuracy of the Gaia $\vlos$ estimates),
\item proper motion determination errors, $\sigma_{\pm}$, which would appear as distance over-estimates, but are again negligible by more than an order of magnitude,
\item the tilt of the velocity ellipsoid, which is of some importance for the statistics. This term is important, as the radially elongated velocity ellipsoid produces a locally changing correlation between the heliocentric velocity components, which can partially line-up with the $\Tuw$ and $\Tvw$ angle combinations.
\end{itemize}
We have to add the systematic uncertainties from these terms to our error budget, assuming that the uncertainties in these terms are statistically independent of each other. As already done in previous studies \citep[][]{SBA} we assume a systematic uncertainty of $10 \%$ of the calculated correction value for the first two terms, and an uncertainty of $30 \%$ for the turn of the velocity ellipsoid correction.

As stated above, due to the unprecedented precision of Gaia, the exact values of proper motion errors do not matter here as long as the order of magnitude of the uncertainty estimates in the Gaia pipelines is correct. Similarly, the error correlations are mostly inconsequential: two team members did independent tests on independently calculated mock samples, where we folded the mock measurements with the full error matrix between parallaxes and proper motions as given for each star in the sample. In these tests, the effect of error correlations on our statistics is more than one order of magnitude less than our systematic and formal error budget for whole-sky measurements. On pencil beams, like in tests of high $\eclat$ stars, it contributes of order one tenth to the residual bias (see analysis below).

For the velocity ellipsoid correction, we assume that the velocity ellipsoid points to the Galactic centre at every position. Other than in previous applications of the \cite{SBA} method, the Gaia sample spans a large volume throughout the disc, and when we select by distance, stars in each sample will cover regions with vastly different values of the velocity dispersion. To optimise the estimate for this correction term, we directly measure the velocity dispersions in the Galactocentric spherical coordinate frame weighted by their impact on the distance estimator.

As described in Section~\ref{sec:Generalthought} the method does not assume any velocity ellipsoid, and does not even require knowledge of the correlations between the velocity components (e.g. $\Uh$ and $\Wh$). The only important requirement is that Galactic structure does not infer a correlation between this velocity correlation and Galactic position. Realistic structure (e.g. a stream passing through the survey, or disc breathing modes) might produce a local velocity correlation (and so distance statistics on small patches of sky are uncertain, see the $\eclat$ issue below), but cancels out to first order with large sky coverage. We have already tested and confirmed this on realistic simulations in the Appendix of \cite{SA17}.

\begin{figure}
\epsfig{file=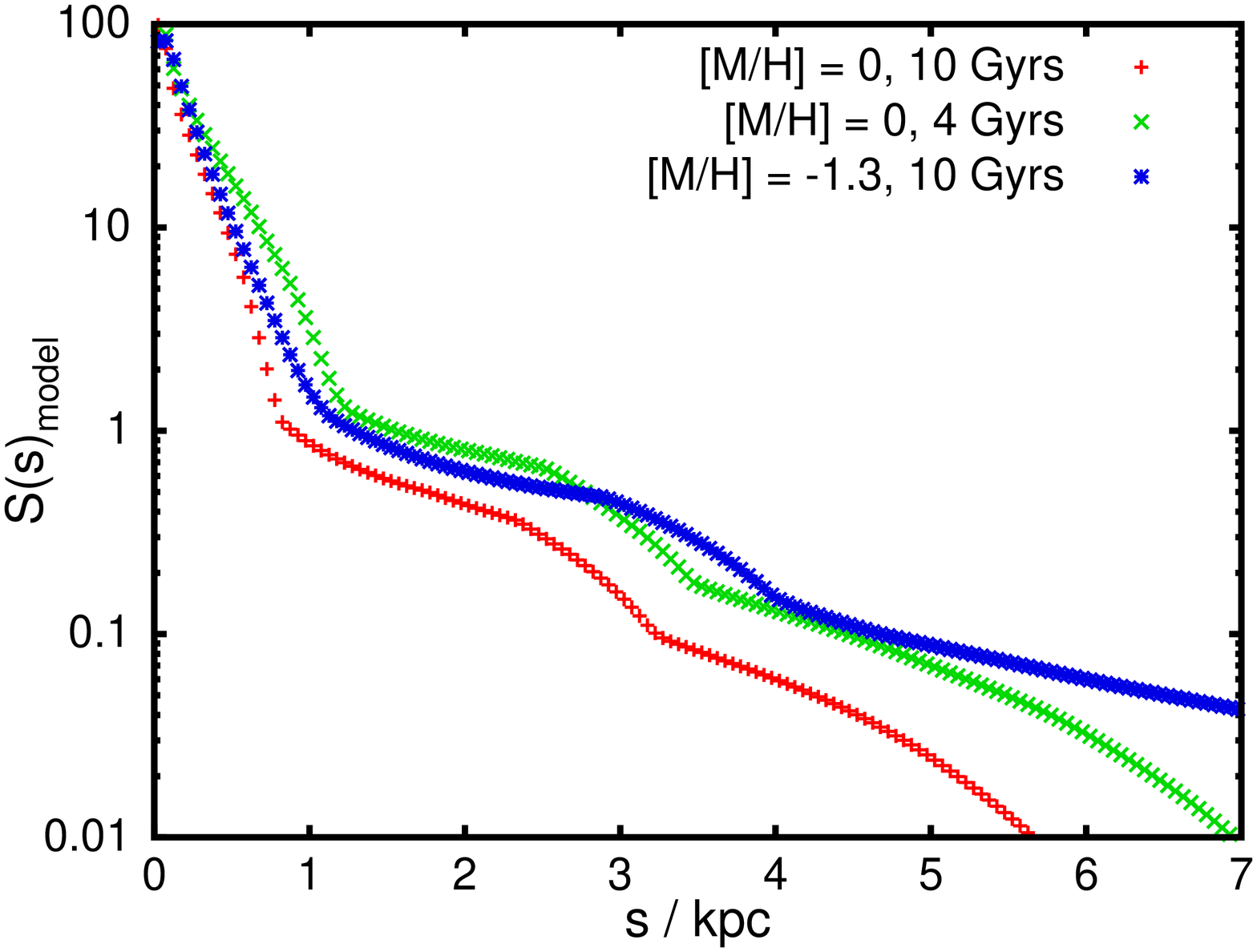,angle=-90,width=\hsize}
\epsfig{file=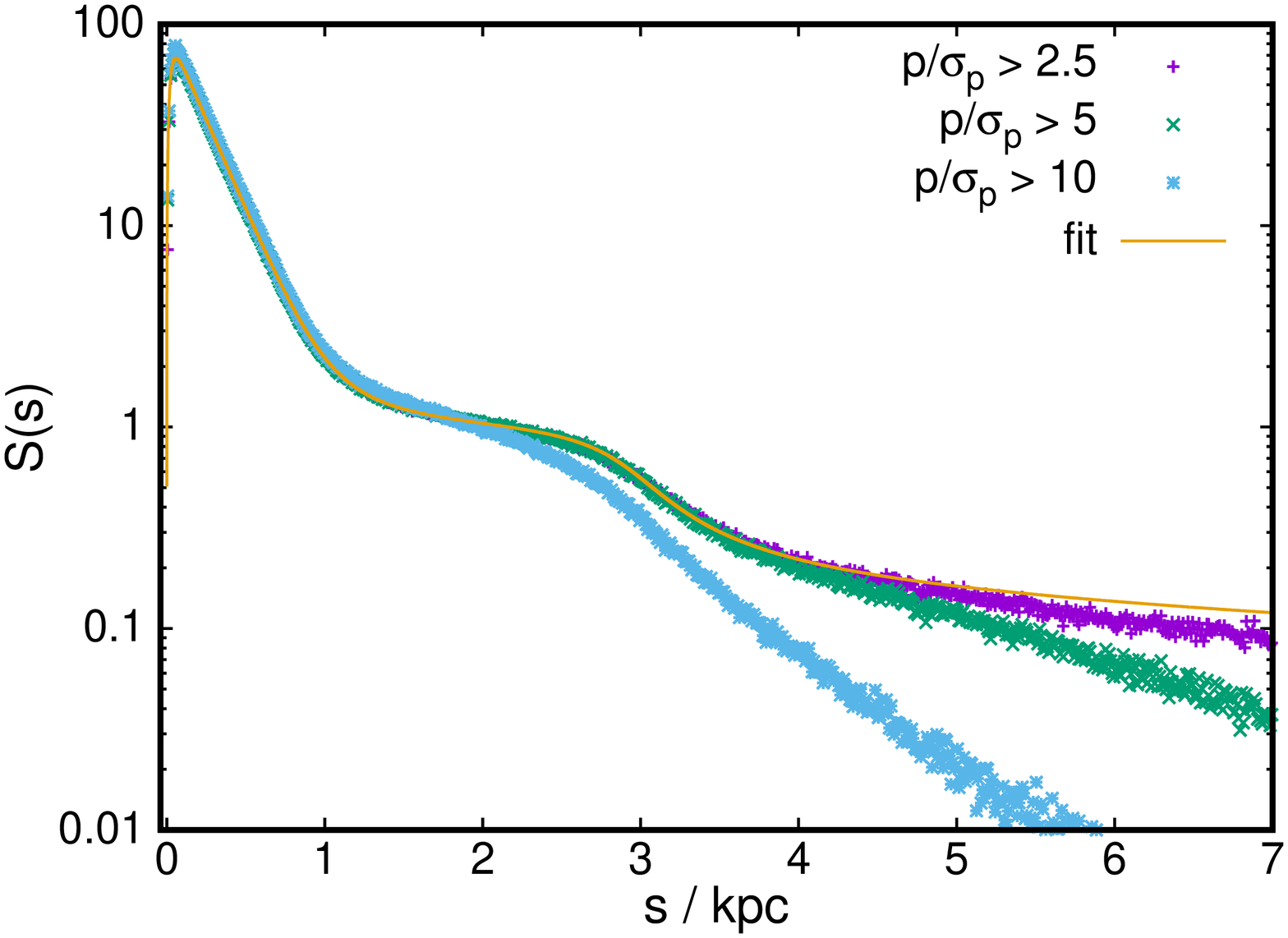,angle=-90,width=\hsize}
\caption{Top panel: selection function $S(s)_{\rm model}$ in distance from a simple population synthesis for populations with fixed metallicity and age and a simplified magnitude-dependent selection function resembling the Gaia DR2 subset with $\vlos$ measurements. The normalisation is irrelevant, so we multiplied with an arbitrary factor to match the bottom panel.
Bottom panel: Selection function measured from the data after several iterations with different quality cuts on the relative parallax quality. The least tight cut ($\vpar / \sigma_p > 2.5$) should not be used in stellar sample selections as it will comprise a large number of catastrophic distance errors. It can, however, serve to educate on the real shape of the distance-dependent selection function $S(s)$. Note that the normalisation is irrelevant and here comprises amongst other effects a geometric factor $2 \pi$.
}\label{fig:priortest}
\end{figure}

\section{Distance Determinations and Selection Function for Gaia DR2}\label{sec:Distances}

Before we can test distances/parallaxes with the SBA method, we have to derive distance expectation values for all stars, using the method of \cite{SA17}. From parallax measurements we calculate the probability distribution in distance $P(s)$ for each star by
\begin{equation}
P(s) = N^{-1} s^2 G(p, p_0, \sigpar)\rho(s(\vpar),\gl,\gb)S(s(\vpar)) $,$
\end{equation}
where
\begin{equation}
N = \int{ds s^2 G(p,p_0,\sigpar)\rho(s(\vpar),\gl,\gb)S(s(\vpar))}
\end{equation}
is the normalization, $s$ is the distance from the Sun, $\vpar$ denotes a parallax, $G(p,p_0,\sigma_p)$ is the (Gaussian) observational likelihood distribution in parallax, given the measurement $p_0$ and effective uncertainty $\sigma_p$, and $\rho(s(p),\gl,\gb)$ is the assumed density model. $S(s)$ denotes the selection function, i.e. number of stars detected in the sample divided by the number of stars actually there. As in previous papers, we use the simple density model from \cite{SB14}, which contains a thin disc, thick disc, and halo component (we can neglect the bulge because we only select stars with $|\gb| > 10 \,\degs$). All calculations are done with a self-adaptive trapezoid integration, where we start from the suspected maximum of the PDF and integrate to both sides in distance simultaneously, adapting the step-length upwards when the relative contribution of each segment to the integrated value falls below a threshold. We tested that our results are the same when lowering the initial step-length, and threshold by a factor $10$, or when using a simple step-wise integrator.

We tested by cutting the sample in Galactic latitude and longitude that this model is sufficiently close to the underlying distribution to ensure a good distance measurement. The most important quantity in the above equation, however, is the distance-dependent selection function $S(s(p))$, which arises from the magnitude-dependent cuts of the sample. Predicting $S(s)$ directly would require a full chemodynamical galaxy model including a three-dimensional reddening map. With this choice, we would be vulnerable to systematic uncertainties of the model choice, stellar evolution models, and the reddening map. 

Here, we instead measure $S(s)$ from the data themselves. Nevertheless, it is useful to formulate a model-based expectation, which acts as a sanity check to our results. The top panel of Fig. ~\ref{fig:priortest} shows the selection function $S'(s)$ calculated from simple population synthesis using the machinery of \cite{SM17} using B.A.S.T.I. stellar models \citep[][]{Piet04, Piet09} and a simple \cite{Salpeter55} Initial Mass Function (IMF). The shown $S'(s)$ expresses the number of stars per solar mass (at birth) of a stellar population at a given distance $s$. Since the normalisation factor here is irrelevant for our purposes, we multiplied $S'(s)$ with an arbitrary factor $80$ to facilitate direct comparison with the measured $S(s)$ in the bottom panel. For this figure, we suppose a simple magnitude-dependent selection, setting the selection probability constant below Johnson V-band $m_V  < 12.8 \magn$ and then going linearly to zero for $m_V \ge 13.5 \magn$. 

$S(s)$ has three main regions: i) a quite steep roughly exponential decrease of $S'(s)$ with distance $s$ in the near field, stemming from the magnitude limit moving up through the main sequence and turn-off region with increasing distance modulus. The exponential behaviour is expected, since the magnitude scale is logarithmic, the luminosity of stars is roughly the fourth power of the mass, and the IMF is close to a simple power law (as we can neglect the impact of stars below $\sim$$0.5 M_\odot$ where this is no longer true). This decrease is followed by ii) a weakly inclined plateau associated with the subgiant branches and red clump, followed by a downward knee and somewhat shallower decline at the largest distances due to iii) the red/asymptotic giant branch stars. There are some differences between different ages and metallicities: metal-poor stars are somewhat more luminous on average, shifting the features to the right, older populations have their subgiant branch at fainter magnitudes and less stars overall on the giant branch (the lengthening of stellar main sequence lifetimes outweighs the increase in the IMF). However, the common features let us predict the functional shape for the selection function rather well, when averaged over all stellar populations. We choose:

\begin{equation}\label{eq:selfunc}
S(s) = a A(s) B(s) C(s)
\end{equation} 
where $a$ is a normalization constant, and the three multipliers are:
\begin{eqnarray}
A(s) &=& \exp{(-bs)} + c\exp{(-ds)}\frac{1.0}{1.0 + \exp{-h(s-j)}} \nonumber \\
B(s) &=& \frac{1.0}{0.5\pi + k_2}(\tan^{-1}{(l(l_2 - s))} + k_2) \nonumber \\
C(s) &=& 1 - \exp{(-zs)} \nonumber
\end{eqnarray}

The rationale behind this is to capture in $A(s)$ the general shape with two exponentials of scale-length $b^{-1}$ and $d^{-1}$ for the short (out to $\sim 1 \kpc$) and long (beyond $3 \kpc$), model the step in $S(s)$ with $\tan^{-1}$ in $B(s)$, as well as the (unimportant) drop-out of luminous or otherwise too close stars (potential loss due to proper motions) with $C(s)$. 

We proceed as in \cite{SA17}, iterating the distance calculation with the adapted prior. However, due to the good nominal quality of Gaia parallaxes, the general appearance of the measured $S(s)$ is already present from the first iteration using a flat prior. A fit of the function to data corrected for a mean parallax offset of $\delpar = -0.048 \mas$ (see following below) after several iterations of the prior fitting is presented in Fig. \ref{fig:priortest}, the values are provided in Table \ref{tab:selection}. The shape matches the prediction from the population synthesis. We also checked that the selection function does not vary strongly with galactic latitude, which signals that the spatial prior is sufficient and population differences throughout the galaxy will not strongly bias our distance estimates.

\begin{table}
\caption{Parameters of the selection function (eq.~\ref{eq:selfunc})
}\label{tab:selection}
\begin{tabular}{lll}
Parameter & value & unit\\ 
\hline
$a$ & $ 97.9594$ &  \\
$b$ & $ 4.10228$ & $\kpc^{-1}$\\ 
$c$ & $ 0.0159642$\\ 
$d$ & $ 0.15$ & $\kpc^{-1}$\\ 
$l$ & $ 2.22279$ & $\kpc^{-1}$\\ 
$l_2$ & $ 2.97547$ & $\kpc$ \\ 
$k_2$ & $ 2.0704$\\ 
$j$ & $ 0.956364$ & $\kpc$\\ 
$h$ & $ 4.91193$ &  \\
$z$ & $ 0.024$ & $\kpc^{-1}$ \\
\hline
\end{tabular}
\end{table}

\begin{figure}
\epsfig{file=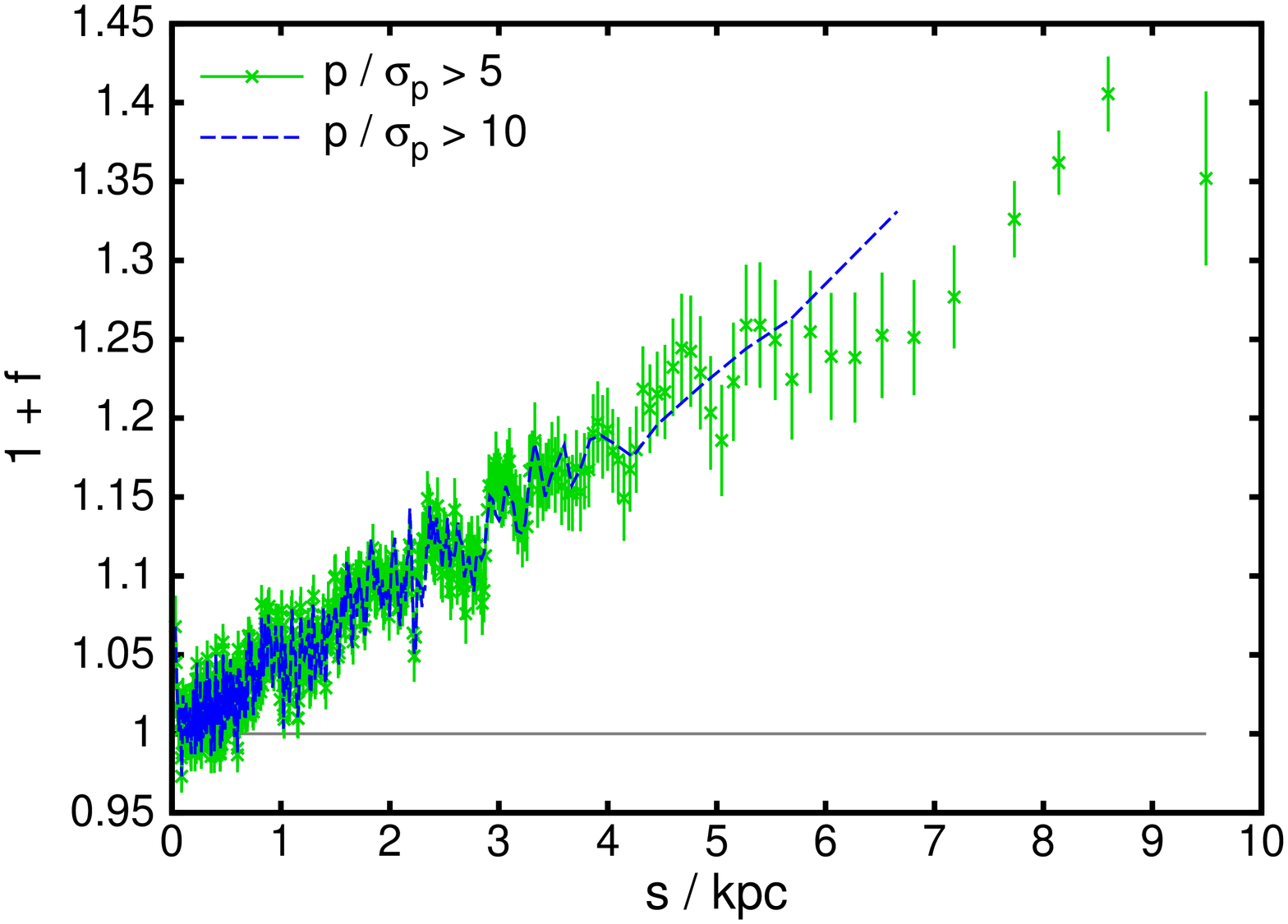,angle=-90,width=\hsize}
\epsfig{file=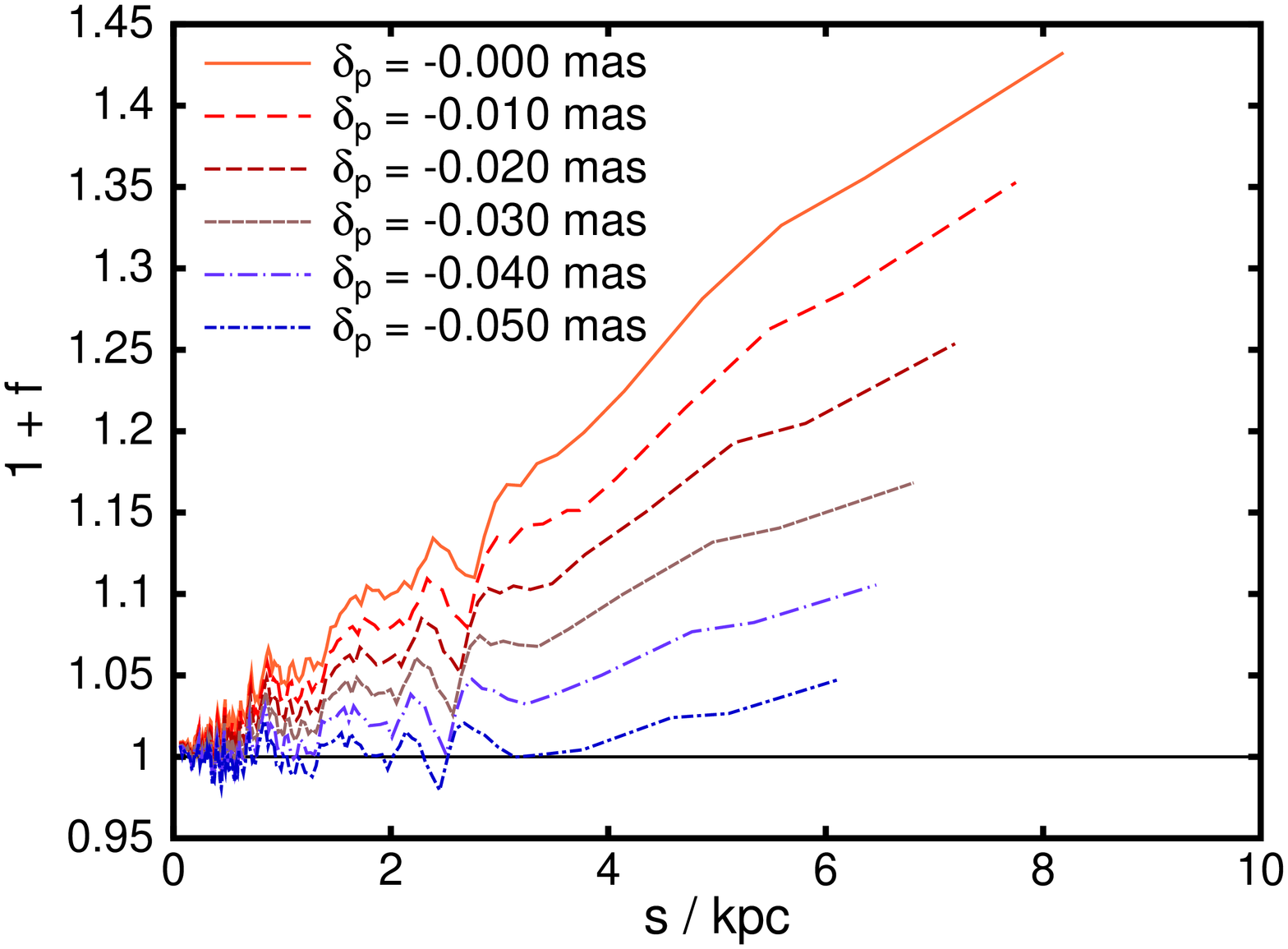,angle=-90,width=\hsize}
\caption{Relative distance error $1+f$ in the Gaia DR2 RV sample when binning the sample in distance. In each panel, we sort the sample in distance $s$ and then let a mask of $15000$ stars width slide by steps of $5000$, i.e. every third data point is independent. The $1+f$ shown on the y-axis, measured with the SBA method, denotes the factor by which the average distance in each sub-sample is wrong. E.g. $1+f = 1.1$ means that stellar distances are on average $10\%$ too large. The top panel shows that the result does not depend on the value of the quality cut in the relative parallax error $\vpar / \sigpar$. The bottom panel (using a sample size of $90000$ in steps of $30000$) uses $\vpar / \sigpar > 5$ and varies a fixed parallax offset, i.e. incrementing all measured parallaxes by $\delpar$, which reduces the estimated distances.}\label{fig:distscan}
\end{figure}

Apart from $d$, which sets the scalelength of the long exponential component, all parameters are well-constrained in the fit. The latter suffers from the fact that we cannot fit beyond the point where the parallax quality cut affects the sample. This cut must not enter $S(s)$, since the stars are culled after they have passed the magnitude limits. Inspection of Fig. \ref{fig:priortest} shows that given the relative drop-out rates of stars at different quality cuts, the fit must be close to the real shape. Extensive further tests of the far-distance end favoured an additional flattening of $S(s)$ by multiplying with $\exp(-(s'/\kpc - 4)/0.07)$, where $s' = {\rm min}(s,10\kpc)$. The difference can be inspected between the top panel of Fig.~\ref{fig:distscan}, which does not contain this factor and the bottom panel, which does include it. It mostly serves to improve a slight kink in the distance statistics. The latter factor only has a minor effect on very remote stars beyond $s > 3\kpc$, and we will apply it to all data shown from Fig. \ref{fig:quantoff} onward. Both the model expectations and the derived distance statistics for very remote stars, which react strongly to the prior, support this choice. 

\section{Distance bias vs. distance: evidence for the parallax offset}\label{sec:Measurement}

\subsection{Scanning vs. distance: Detection of the parallax offset}

Fig. \ref{fig:distscan} measures the relative distance bias $1+f$ when using the \cite{SBA} method for the stellar sample when ordered and binned in the measured distance $s$. The top panel shows this scan of $1+f$ vs. $s$ for two different quality cuts in the relative parallax error, allowing a maximum $\sigpar/\vpar$ as given by Gaia DR2 of $10 \%$ or $20 \%$ respectively. It is apparent that $1+f$ increases nearly linearly with $s$, reaching a relative distance bias in excess of $5 \%$ at $s \sim 1 \kpc$ and errors in excess of $25 \%$ near $s \sim 5 \kpc$. Such a bias precludes a precision measurement of galactic kinematics even in the relative near field. The comparison of different $\sigpar/\vpar$ cuts proves that this cannot be an issue with the distance priors or other parts our measurement method. Otherwise, a tighter quality cut would drastically reduce the bias $1+f$ at a given distance. This leaves only one culprit: a bias in the Gaia parallaxes. 

In fact, as we can see from the bottom panel in Fig. \ref{fig:distscan}, this linear trend is a signature imprint of such a parallax bias. When we apply the correction $\delpar$ to all parallax measurements, we can minimise the trend for $\delpar = -0.048 \mas$ with an uncertainty of about $0.006 \mas$. The most distant bins beyond $s > 3 \kpc$ are in line with this estimate within the uncertainties. Small differences at these distances should be ascribed to uncertainties in $S(s)$ as discussed above. The offset is about double the amount found by \cite{Lindegren18}, but exactly in line with the asteroseismic evaluation of Gaia DR2 \citep[][]{Zinn18}. We note again that we do not believe the evaluations in \cite{Arenou18} and \cite{Lindegren18} on quasars to be applicable to our case, since those are in different apparent (magnitude) window classes with separate astrometric calibrations, and have all different kinematics/zero intrinsic parallax.

We further note that there is a significant positive bias $f \sim 5\%$ for the distances of the nearest stars or, equivalently, brightest parts of the sample (seen in the left-most green datapoint in the top panel Fig.~\ref{fig:distscan}). This bias is near impossible to explain with bad $\vlos$ measurements, which would feign a negative $f$. This bias is for these nearby stars orders of magnitudes larger than the previously discussed parallax offset, which for these stars is negligible. Some of this may be traced back to the magnitude dependent deviations (see below, Fig. 9), some to mis-identifications in the near field.

\begin{figure}
\epsfig{file=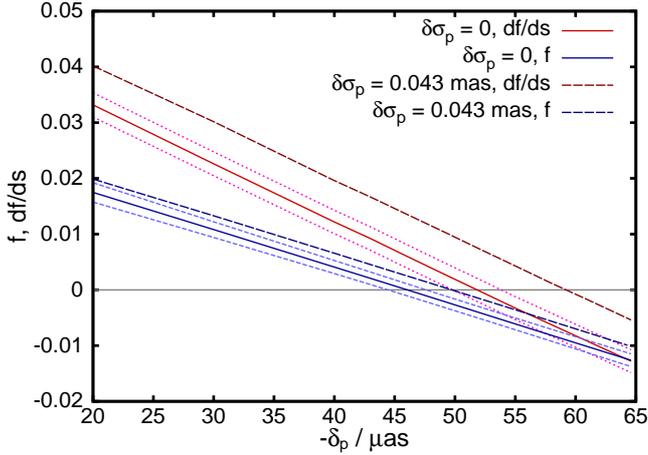,angle=-90,width=\hsize}
\caption{A quantification of the parallax bias $\delpar$ in the sample. Here, we evaluate $f(s)$ for different values as in Fig.\ref{fig:distscan}. These values we fit in the interval $0.1 < s/\kpc < 3$ both with a constant value, and with a linear regression $f(x) = a + \frac{df}{ds}s$ and show the results with the red and blue lines (including their formal $1 \sigma$ error intervals depicted with short-dashed lines). For comparison, we show the same evaluation, but now increasing the parallax error $\sigpar = \sqrt{\sigparg^2 + (\delta\sigpar)^2}$ by adding $\delta\sigpar = 0.043 \mas$ in quadrature (long-dashed without error bars).}\label{fig:quantoff}
\end{figure}

So far, we simply applied a parallax offset. However, from \cite{Lindegren18} a position dependent variation in the parallax bias is expected. There has been discussion in the Gaia collaboration, if such a bias should be added to the error budget or not. Now, in our case, we are interested in the uncertainty for each single star. A priori, we do expect a random, but spatially correlated fluctuation of the parallax offset to enter single star parallax uncertainties just like an additional term that has to be added in quadrature to the formal uncertainty given by the pipeline, setting $\sigpar = \sqrt{\sigparg^2 + (\delta\sigpar)^2}$. 

We thus quantify the correction $\delpar$ on two versions of the sample, once with and once without systematic $\delta\sigpar$, which we take to be $0.043 \mas$ following the quasar analysis of \citet[table 4]{Lindegren18}. One can argue for two ways of measuring the parallax bias: On one hand, we could demand that the average $f$ in the safe region $0.1 < s/\kpc < 3$ should be exactly $0$. While we have high confidence in the accuracy of the distance statistics in this area, one might still want to get rid of the need for a correct zero point. Indeed, since a parallax offset gives an almost linear $f(s)$ dependence, we can alternatively demand that the estimated slope $df/ds$ should be zero. However, within our systematic uncertainties, both methods, shown in Fig.~\ref{fig:quantoff} yield the same values for $\delpar$. Drawing a mean estimate from the shown results and further tests of varying quality cuts, we conclude that $\delpar = -0.048 \mas$ with a negligible formal uncertainty and a systematic uncertainty of $\sim 0.006 \mas$. The systematic uncertainty is a cautious estimate. The systematic uncertainties on velocity ellipsoid and measurement uncertainties are already priced into the formal errors, but we price in a correlated systematic error between evaluation bins, and performed a variation of fit parameters (region in $s$ to fit on), changes in quality cuts (see below), and comparison of the different methods. These would advise a mildly smaller number, and we added a budget for effects that we may have missed. Separate from these effects, accounting for $\delta\sigpar = 0.043 \mas$ further increases the $|\delpar|$ estimate by about $0.006 \mas$ to $\delpar = -0.054 \mas$. The small difference between the $f$ and the $df/ds$ estimation for $\delpar$ can be ascribed to bad luck as it is within $\sim 2 \sigma$. We think, however, that it has contributions by one or several of the secondary problems identified below.

\begin{figure}
\epsfig{file=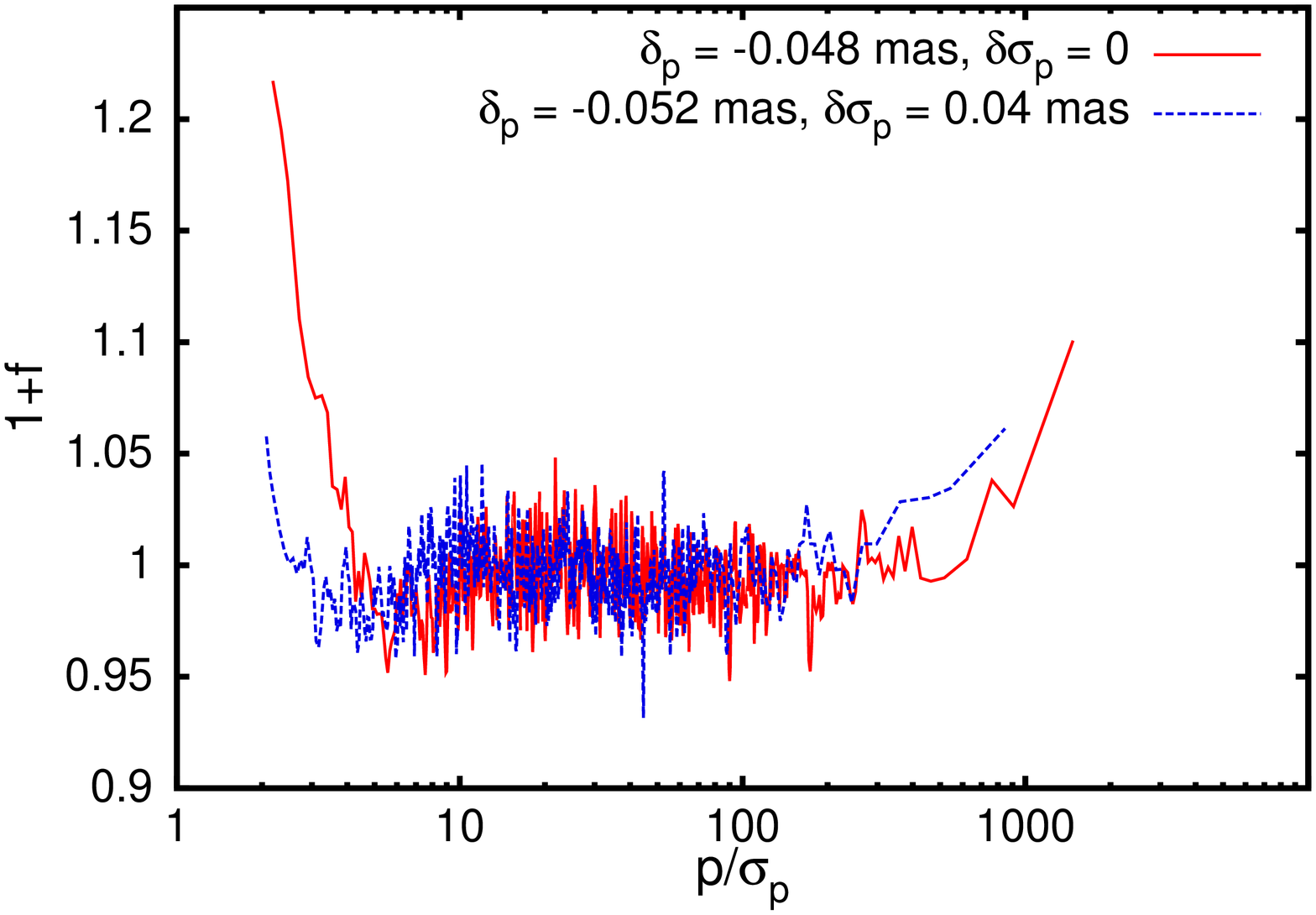,angle=-90,width=\hsize}
\epsfig{file=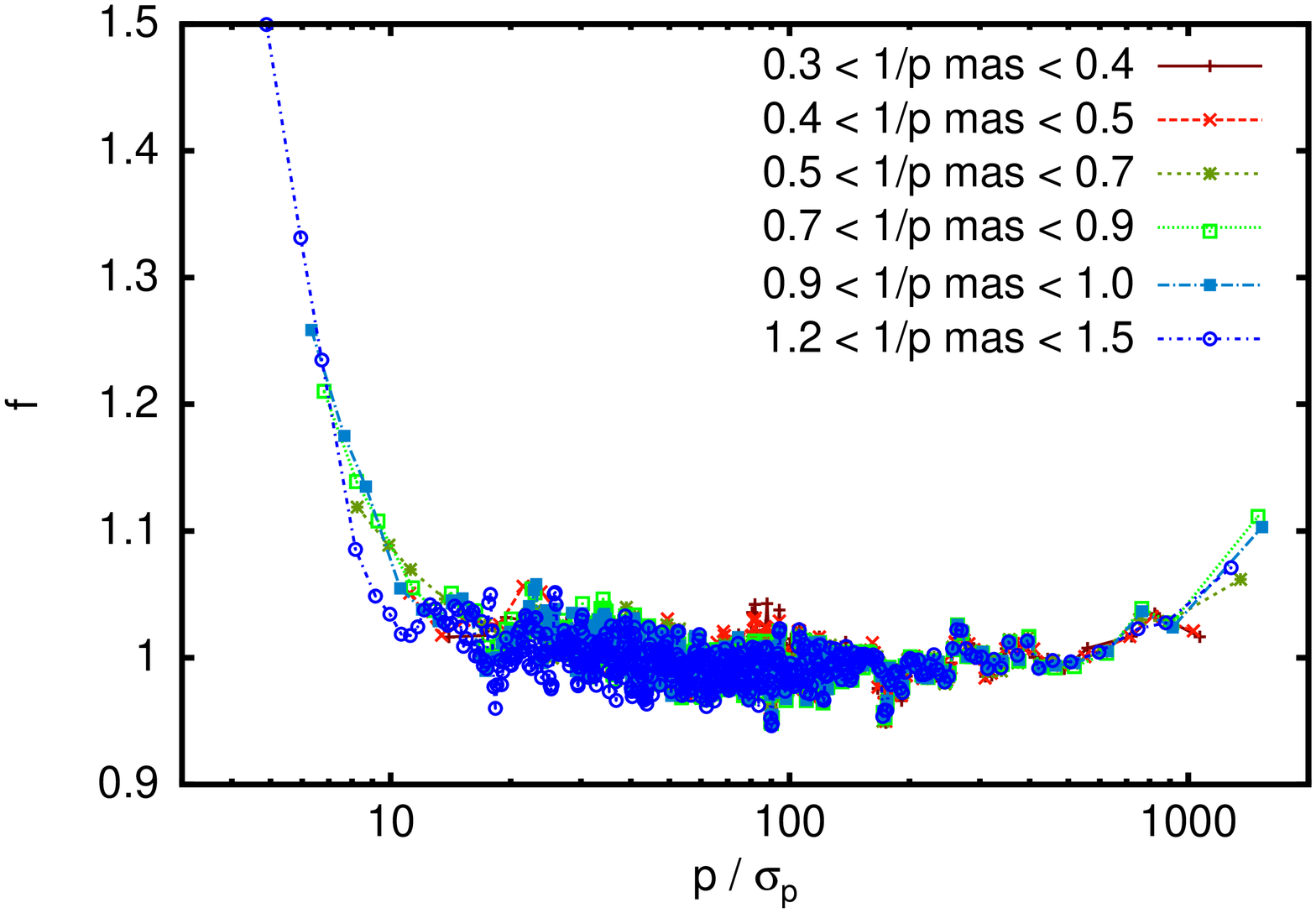,angle=-90,width=\hsize}
\caption{Relative distance error in the Gaia DR2 RV sample when binning the sample in parallax quality $\vpar / \sigpar$. We use the same binning scheme as in the previous figures. In addition we remove all stars with $s,\vpar^{-1} > 10 \kpc$ from the sample. Again the y-axis shows the distance bias factor measured with the SBA method. Both panels reveal a strong increase.  The top panel shows the two possibilities of adding an additional parallax measurement uncertainty of $0.04 \mas$ in quadrature vs. not adding it, the bottom panel displays the same statistics for different cuts in the maximum value of $\vpar^{-1}$.}\label{fig:errorbias}
\end{figure}

\begin{figure}
\epsfig{file=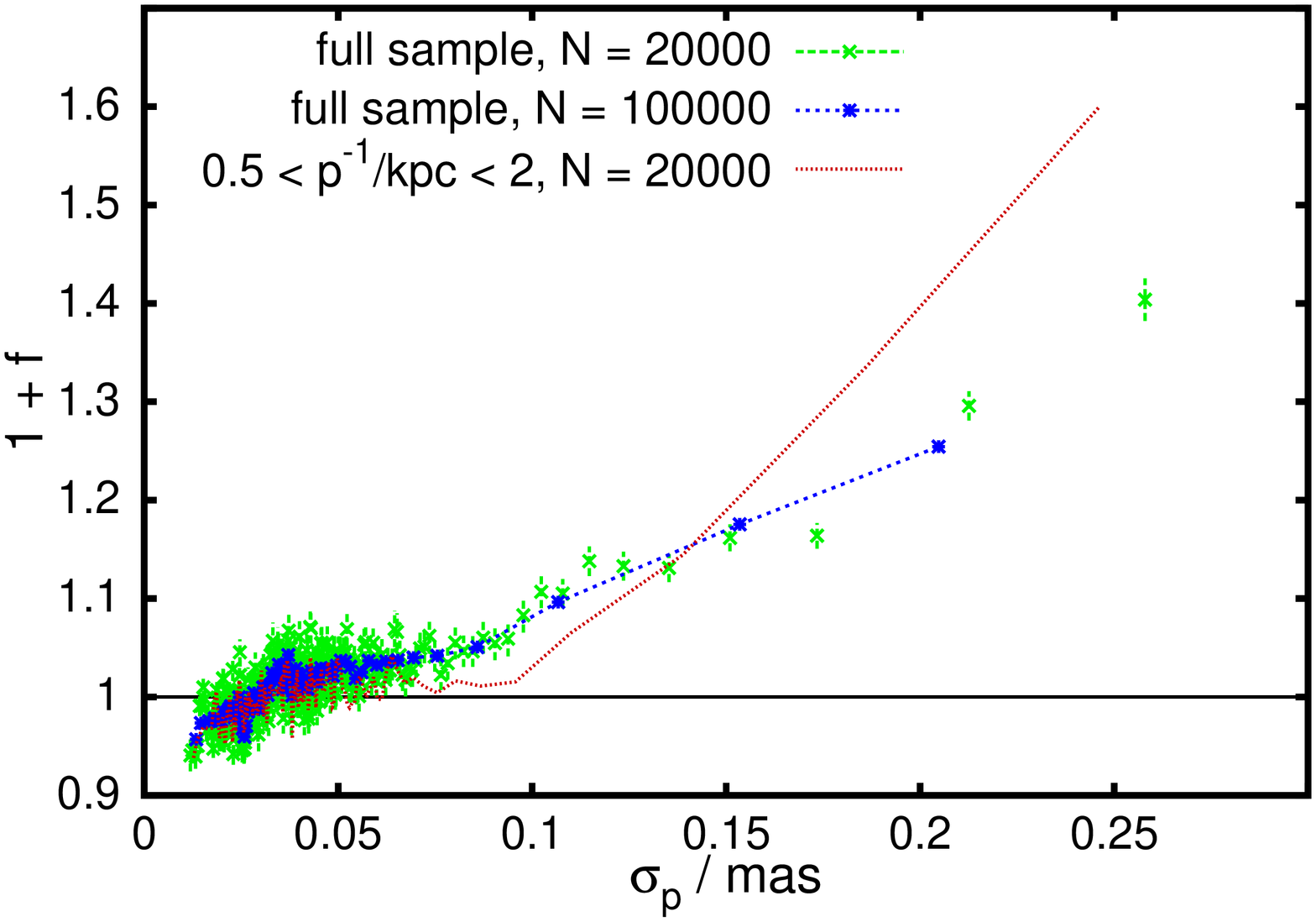,angle=-90,width=\hsize}
\epsfig{file=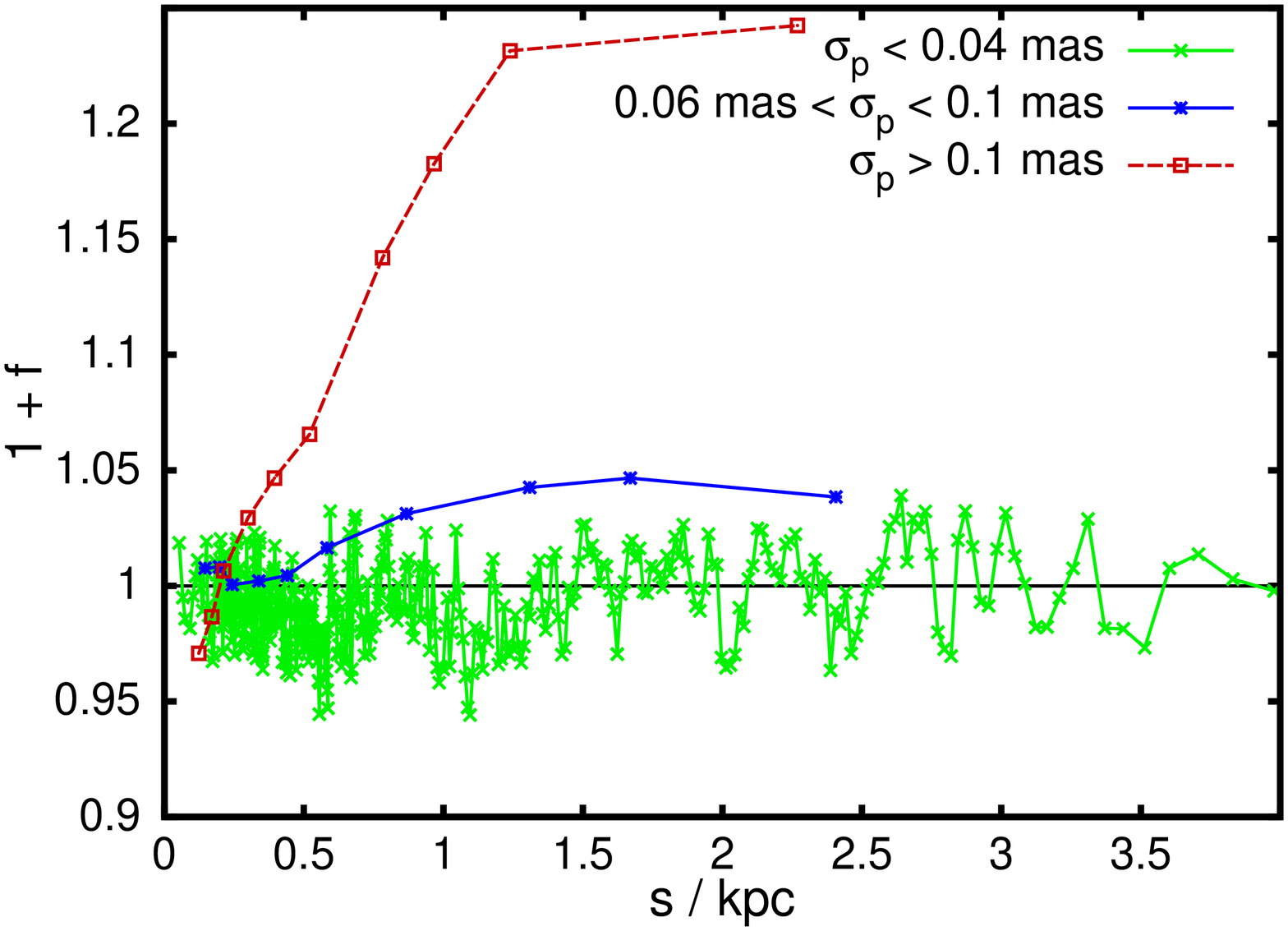,angle=-90,width=\hsize}
\caption{Top panel: Distance bias versus the the parallax error $\sigma_p$ as provided in the Gaia DR2 dataset. The bottom panel displays the distance bias vs. distance to uncover the nature of the trends/biases observed in the top panel. The subsample with large parallax uncertainty, $\sigma_p > 0.1$, shows the typical signature of a larger parallax offset, i.e. the distance bias rises almost linearly with distance. The other subsamples have a weak indication of the same trend.}\label{fig:parerrtrend}
\end{figure}

\begin{figure}
\epsfig{file=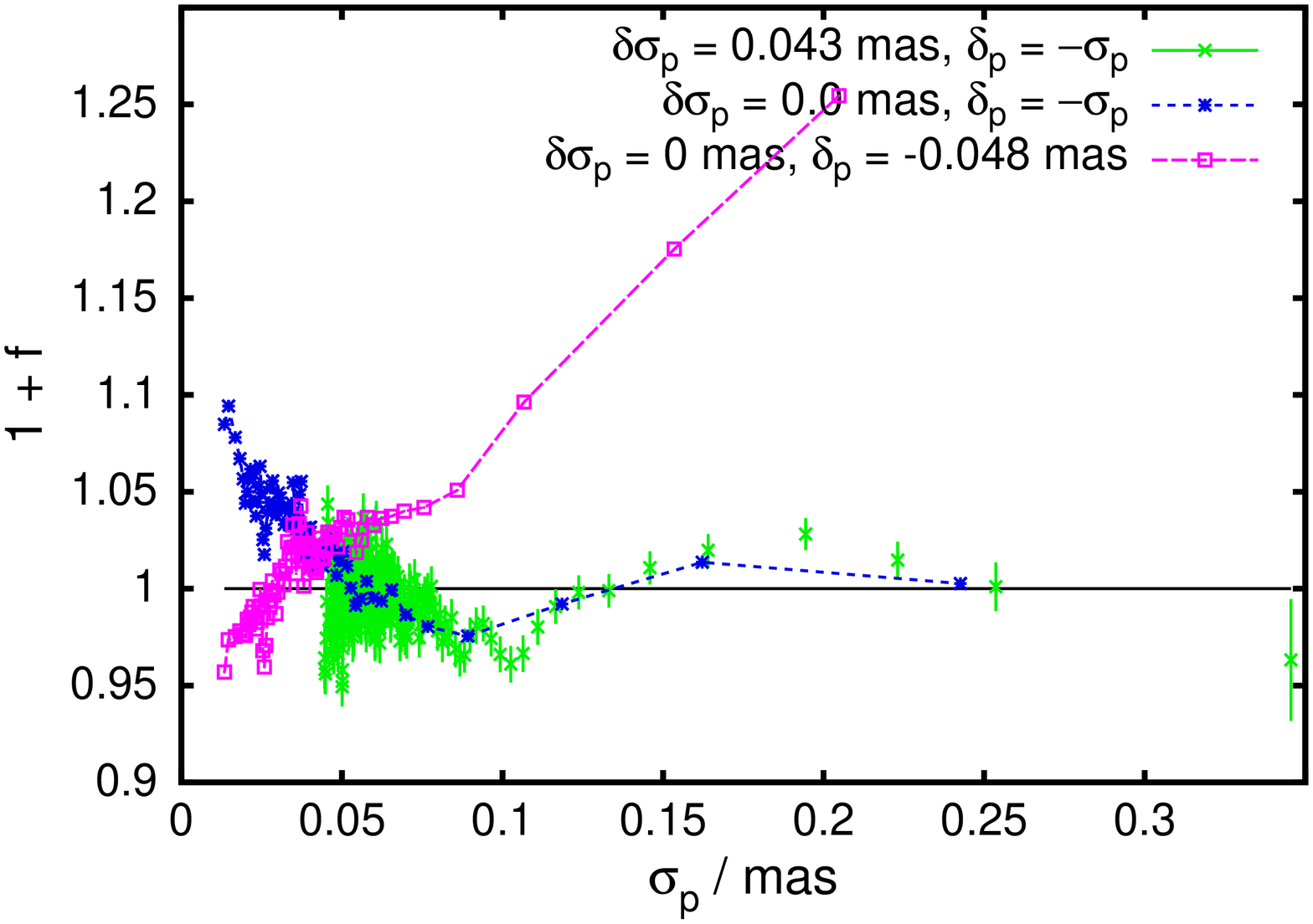,angle=-90,width=\hsize}
\epsfig{file=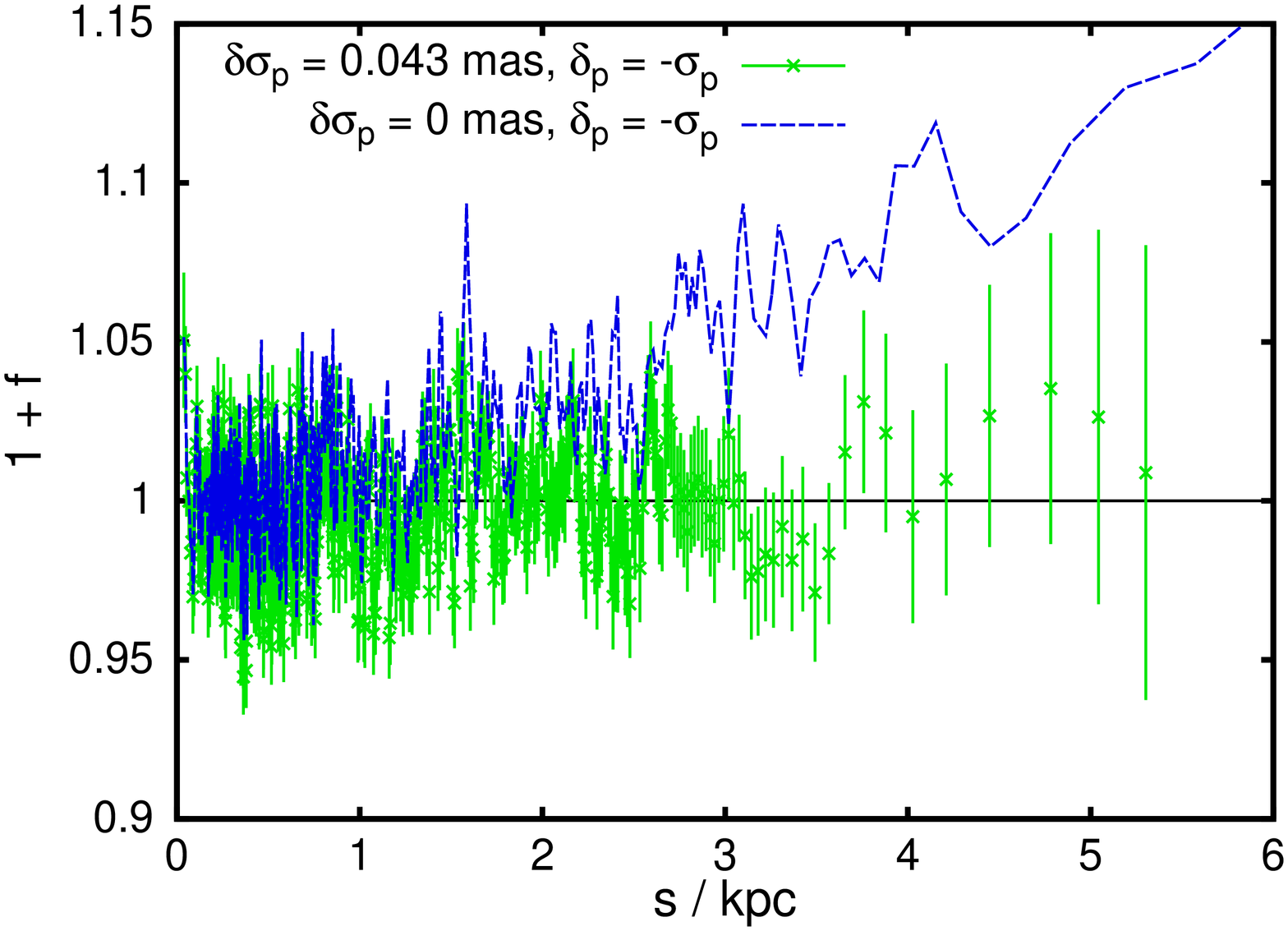,angle=-90,width=\hsize}
\caption{Studying the hypothesis that the parallax error predicts the parallax offset. The top panel shows a scan of distance bias vs. parallax error when assuming that the parallax offset equals the parallax error, i.e. $\delpar = -\sigma_p$, and attempting to correct the error by adding $\sigma_p$ to the parallax before evaluating the distance. The two versions of doing this are shown with green errorbars (adding $\delta \sigma_p = 0.043 \mas$ in quadrature to $\sigma_p$ and the blue line (not adding a parallax error). It is striking, how well the dataset is constructed when assuming this, instead of the constant parallax offset discussed previously (and shown for comparison with the purple line). In the bottom panel, we show a scan in distance of the sample, again showing that the method of first adding $\delta \sigma_p = 0.043 \mas$ in quadrature produces a near-perfect outcome.}\label{fig:parerrsolve}
\end{figure}

\subsection{Quantifying parallax offset vs. uncertainty}

Could we make different assumptions for parallax error visible with our distance method? In principle yes - the background is that the expectation value of a single star's distance shifts systematically with the assumed error. Consequently, a misjudgement of uncertainties will typically show as a distance bias. However, as we saw in Fig. \ref{fig:quantoff}, this effect is subordinate to the other problems in the sample. Nevertheless, assessing the right value for the $\vpar/\sigpar$ quality cut requires a sample scan in this quantity. This is done in the top panel of Fig. \ref{fig:errorbias}. We remind that apart from stars at very small distances (corresponding to large $\vpar/\sigpar$) in this plot, the distance scan (see Fig. \ref{fig:distscan}) shows no significant deviations after correcting for $\delpar$. Yet, the red line in the top panel of Fig.~\ref{fig:errorbias} reveals an impressive increase in $1+f$ towards lower $\vpar/\sigpar$. Note, however, that we have extended the sample to stars beyond our usual quality limit, admitting all stars with $\vpar / \sigpar > 1$; below $\vpar / \sigpar < 4$ there will definitely be a sizeable fraction of stars with catastrophic distance misestimates. Also, the large parallax uncertainty allows for major parts of the probability distribution $P(s)$ to cover highly uncertain regions of $S(s)$. However, extensive experimentation with $S(s)$ could not rectify the abnormalities in Fig.~\ref{fig:errorbias}. One could now be tempted to argue that the suggested inclusion of  $\delta \sigpar = 0.043 \mas$ removes the problem, but this is a deception caused by the left-shift of the graph, since we increased $\sigpar$. The $S(s)$ uncertainty can be resolved by limiting the $\vpar^{-1}$ range to contain the PDFs of even uncertain parallax measurements within the safe region $s < 3.5 \kpc$. The result of this is shown in the bottom panel of Fig.~\ref{fig:errorbias}: there always remains a spike in $1+f$ on the left-hand side. Something else is going on here.

We now restore the parallax quality limit and plot $1+f$ against the pipeline parallax error $\sigpar$ itself in the top panel of Fig.~\ref{fig:parerrtrend}. For $\sigpar > 0.12 \mas$ distance bias sky-rockets, even for the rather conservative cut $\vpar^{-1} < 1.5 \kpc$ applied here. This suddenly clarifies the tension in the plots of $1+f$ against $\vpar/\sigpar$, because we were just moving the region of large parallax errors via the distance/parallax cut. A further, unsettling observation is the uptrend of the distance bias with parallax error between $\sigpar = 0.045 \mas$ and $\sigpar = 0.06 \mas$, turning from a negative to a positive distance bias of order $2\%$. 

The bottom panel of Fig.~\ref{fig:parerrtrend} attempts to qualify the nature of this failure. One could argue that stars with large parallax uncertainties noted by the pipeline should be binaries, affecting their proper motions, and thus our distance statistics. However, we see a clear increase of the distance bias with distance, putting the blame at a parallax offset exceeding by far the $\sim 0.05 \mas$ of the entire sample, for which we already correct in the shown data. Given the displayed results, the only viable explanation is that $\delta \vpar$ is at least to some extent proportional to $\sigpar$.

With such an unusual finding, it is natural to point the search for an honest error at our own code. E.g. there could be a typo in our distance integral creating the dependence on $\sigpar$. Apart from controlling and testing our code, Fig.\ref{fig:parerrsolve} investigates this by checking different assumptions for the distance error. The purple line displays the original trend as found with a constant $\delpar = -0.048 \mas$. The blue line corrects this to setting $\delpar = \sigpar$ and the green line in addition adds $\delta\sigpar = 0.043 \mas$ in quadrature. If we had made a $\sigpar$ dependent error ourselves, the green line and the blue line should deviate in the same way (as the distance estimates depend very weakly on the assumed $\sigpar$). In contrast, we see that only with the full correction and assuming that $\delpar = -\sigpar$, we can rectify the trend in the sample.

\begin{figure}
\epsfig{file=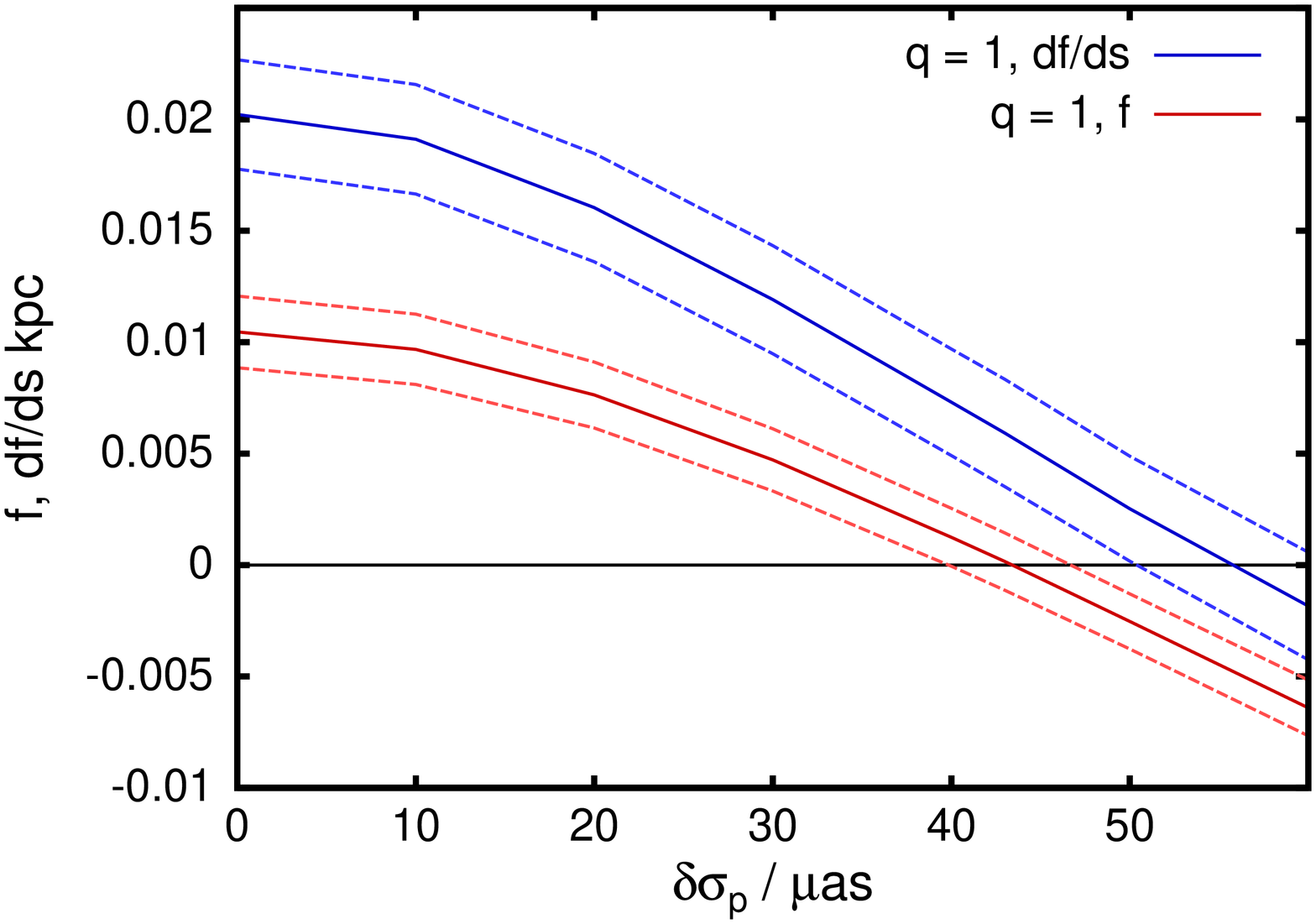,angle=-90,width=\hsize}
\epsfig{file=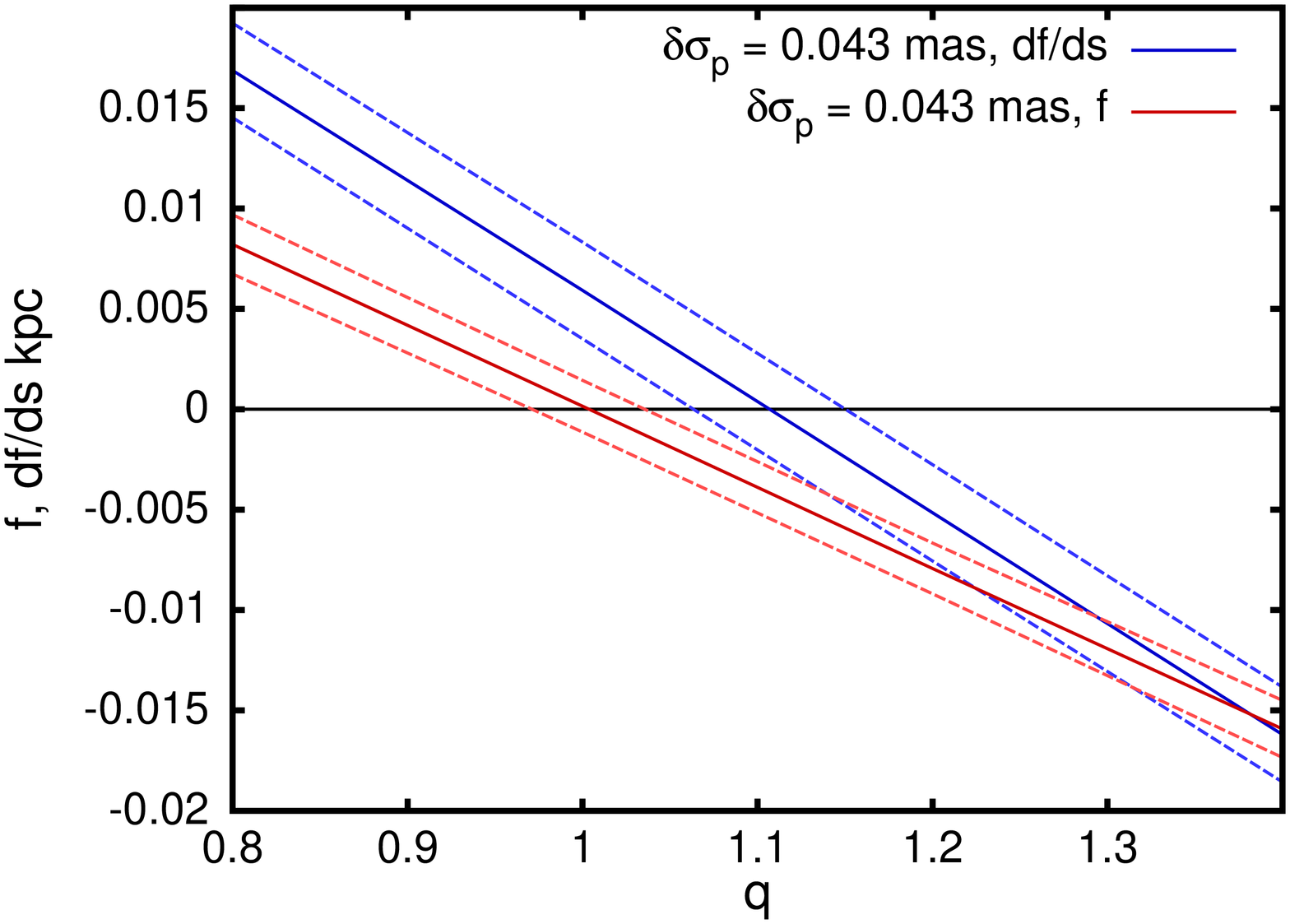,angle=-90,width=\hsize}
\caption{Top panel: Distance bias vs. additional parallax error $\delta \sigpar$. Like in Fig. \ref{fig:quantoff} we bin the sample in distance and measure both the average distance bias (red lines) and the trend of the distance bias with distance for distances smaller than $3 \kpc$. To ensure a clean sample, we removed all stars with $\eclat < -55 \degs$, which, however, has only quite minor impact on our statistics.}\label{fig:quantparofferr}
\end{figure}

While we do not want to adhere to the notion that the parallax offset is perfectly equal to the parallax error, we do in Fig.~\ref{fig:quantparofferr} attempt to show a quantify this dependence. In both panels we use our usual quality cuts in colour, $g$-band magnitude and $\vpar/\sigpar > 4$. In addition, we removed the parts of the sample with $\eclat < -55 \degs$, which, however, has very minor effect. The top panel demonstrates that if we would assume that $\delpar = -\sigpar$, both the average distance error and the trend of distance error with distance are within the systematic uncertainties, in line with assuming the usual additional parallax error $\delta \sigpar \sim 0.043 \mas$ added in quadrature. The bottom panel tests different values of the proportionality constant $q$, when we set $\delpar = -q \sigpar$ after adding the additional term to $\sigpar$. Both statistics are in line with a value very close to $1$.

\begin{figure}
\epsfig{file=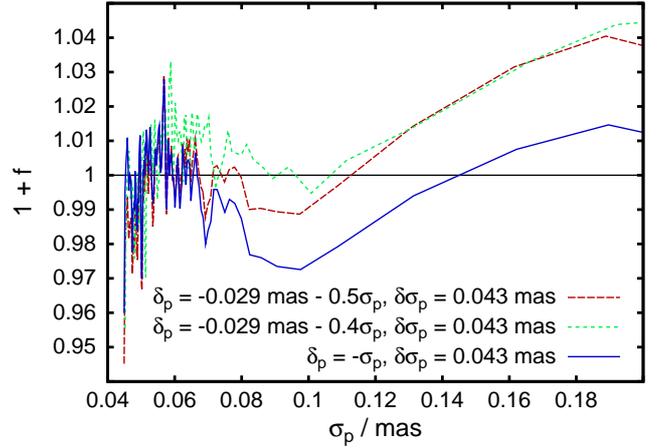,angle=-90,width=\hsize}
\caption{Distance bias $1 + f$ vs. parallax error $\sigpar$. Using the usual quality cuts, we probe different assumptions for changing $\delta \vpar$ and $\sigpar$. The plot uses samples of $90000$ stars, moving the mask in steps of $30000$.}\label{fig:qhalf}
\end{figure}

While the notion of $\delpar = -\sigpar$ is simple at first hand, it is rather unreasonable to believe that there should be such a perfect equality. To this end we tested a third option: While leaving $q$ free again, we added the offset found for the quasars to the parallax, i.e. set $-\delpar = 0.029 \mas + q \sigpar$. Since about half the offset in parallax needed is now again captured by a constant term, we can expect that a good solution will be found for $q \sim 0.5 \mas$. This result is shown in Fig. ~\ref{fig:qhalf}, where we plot the distance bias against $\sigpar$ for both $q = 1$ and $q = 0.5$ plus the constant term. However, various experiments show that we cannot get rid of the trend of $f$ at small $\sigpar$, if $q$ is not close to 1 in this region. The equality is definitely not perfect, since we would require a $q$ closer to $0.5$ in all possible assumptions near $\sigpar \sim 0.09 \mas$ and a larger $q>1$ for $\sigpar > 0.012 \mas$.

To summarize this: The best choice for the sample is to add $\delta \sigpar \sim 0.043 \mas$ in quadrature, and to add $\sim 1$ times the parallax error $\sigpar$ to the parallax. When concerned about precision, we further advise to remove all stars with $\sigpar > 0.08 \mas$, and recommend to separately test stars with $\sigpar \lesssim 0.047 \mas$ for anomalies.

\begin{figure}
\epsfig{file=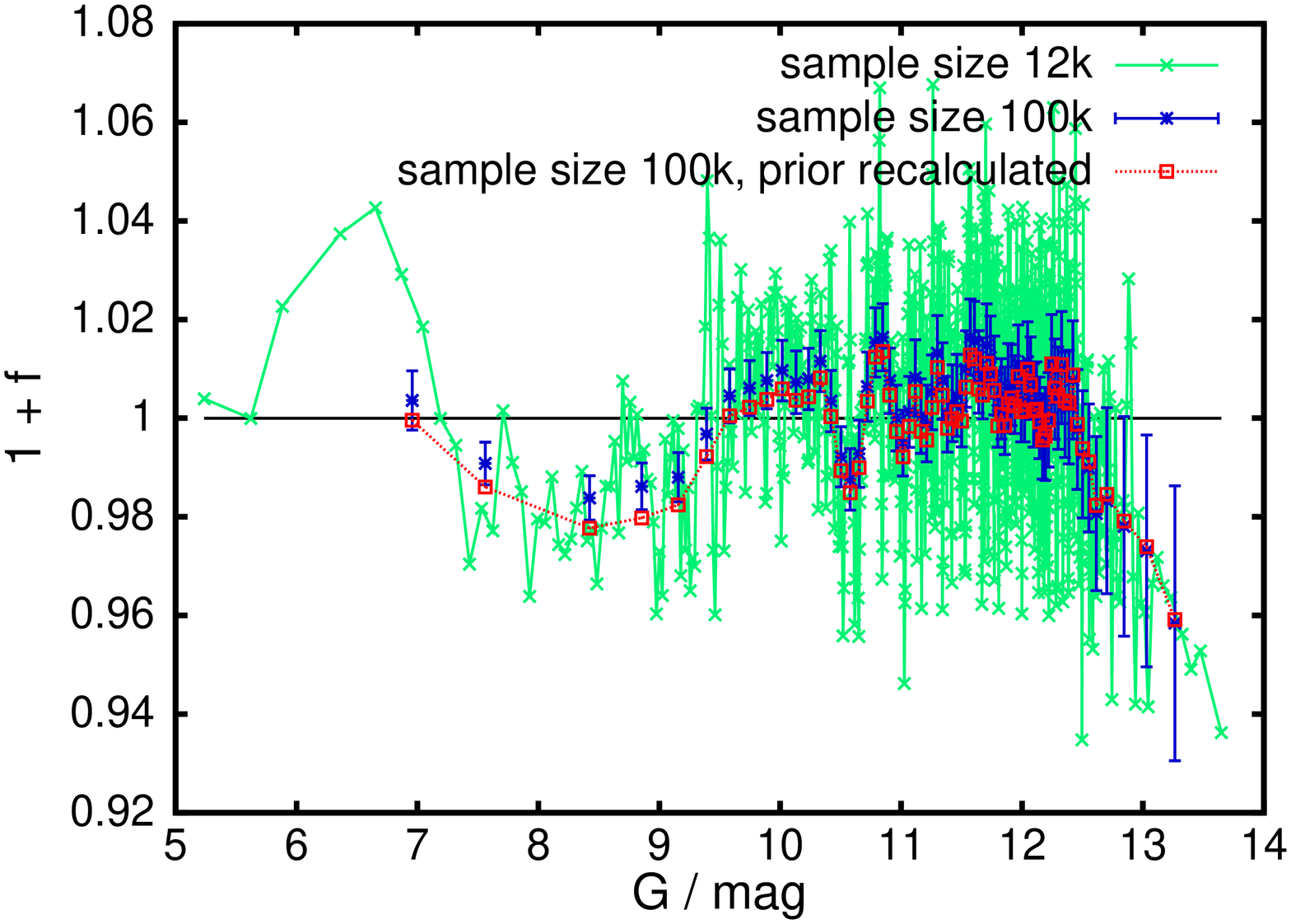,angle=-90,width=\hsize}
\epsfig{file=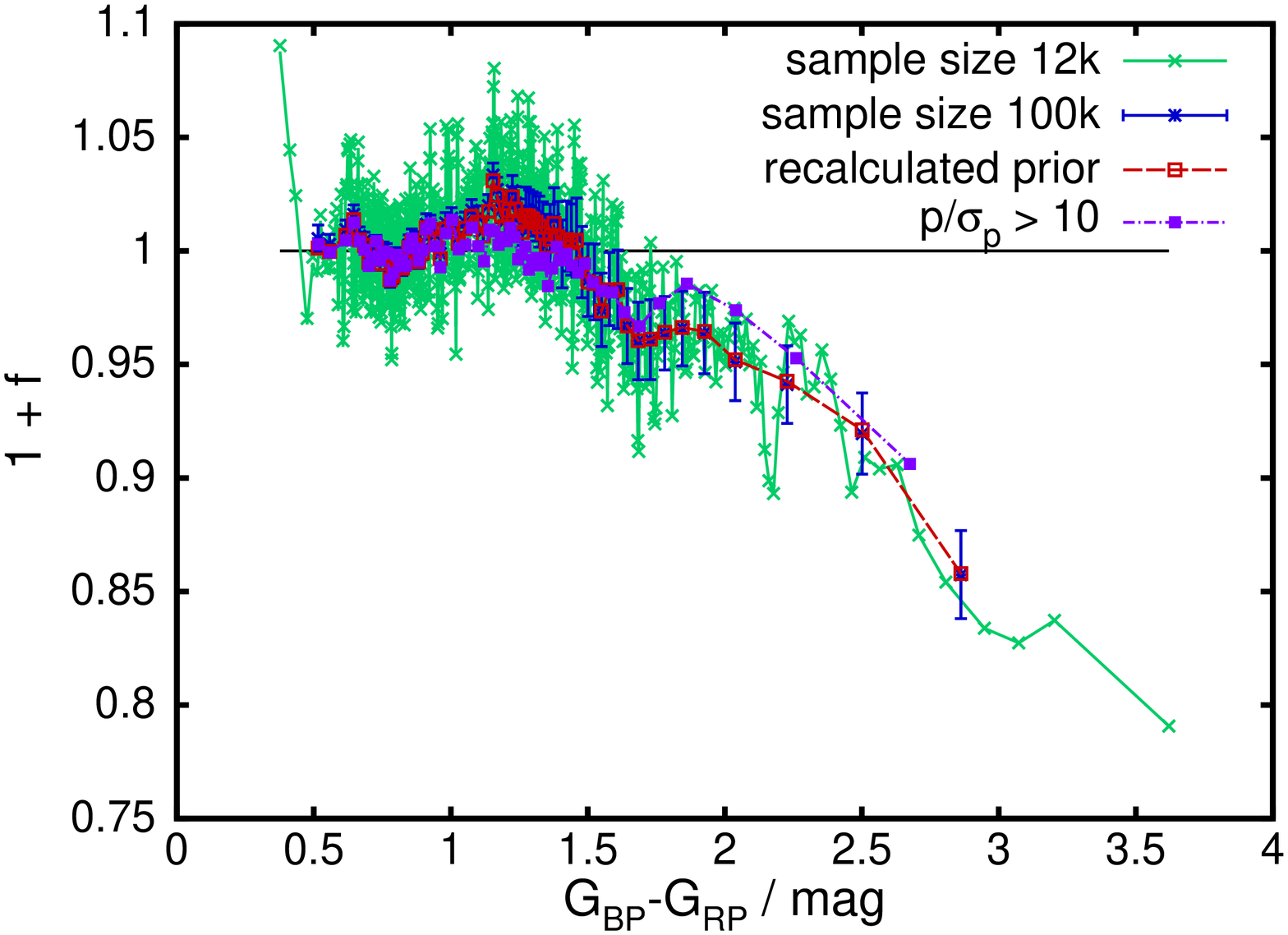,angle=-90,width=\hsize}
\caption{Top panel: Distance bias vs. $\grp$ magnitude of the stars. Here we assume the $\brp-\rrp < 1.5 \magn$ quality cut. The figure displays different sampling sizes ($100000$ moving in steps of $50000$ vs. $12000$ moving in steps of $4000$), and gives one comparison where we recalculated the prior. Bottom panel: Scanning the sample in a similar way in $\brp-\rrp$ colour. Here we assume a $\grp < 14.5 \magn$ quality cut and used the assumption $(\delpar = -0.048 \mas, \delta \sigpar = 0)$.}\label{fig:gmagscan}
\end{figure}

\subsection{Bias vs. colour and magnitude}

After these more complicated considerations, the derivation of safe limits in colour and magnitude for the sample is quite straight-forward. We simply order our sample by $\grp$ or $\rrp$-band magnitude \citep[for an overview of the photometric instrumentation and calibration in Gaia, see][]{Riello18, Evans18}, as well as by the given $\brp-\rrp$ colour, and then iteratively tighten the quality cuts. The results after a first round of clean-ups is shown in Fig.~\ref{fig:gmagscan}. The top panel shows the distance bias vs. $\grp$ magnitude after removal of all stars with $\grp > 14.5 \magn$ and $\brp-\rrp > 1.5 \magn$, while the bottom panel shows a scan in $\brp-\rrp$ colour, after removal of all stars with $\grp > 14.5 \magn$. If we had not censored the faintest stars in the top panel, they would be shown with $1+f \sim 0.8$, i.e. a very strong distance underestimate. Since we recall line-of-sight velocity measurement errors (even if unbiased) look like distance underestimates, the by far most likely explanation for the decline in $1 + f$ is not a failure of Gaia parallaxes, but a much larger than indicated error from the $\vlos$ measurements. If would remain to be explored if the mild decline around $\grp > 13 \magn$ may also be related to a change of magnitude window in the Gaia astrometry. Of course, the change of colour range implies a change in the selection function. We thus devised an automated measurement of the selection function $S(s)$, where we use about $40$ base points for $s < 4 \kpc$, on which we re-measure $S(s)$ and then calculate a grid of relative factors between we interpolate linearly on $\log{s/\kpc}$. As the reader can easily see from the difference between the red and blue points in the top panel of Fig.~\ref{fig:gmagscan}, this more appropriate but far more costly calculation does not significantly change the results. The largest change is around $\grp \sim 8 \magn$, where the re-calculation shortens all distances a little, since $S(s)$ drops more steeply with $s$, and thus exacerbates a little the negative bias in this area. We also note that the relatively sharp and borderline significant feature just below $\grp \sim 11 \magn$ resembles a lot what was shown in the re-calibration comparisons in \cite{Lindegren18}. However, the sample size and small amplitude of the effect precludes further investigation.

\begin{figure}
\epsfig{file=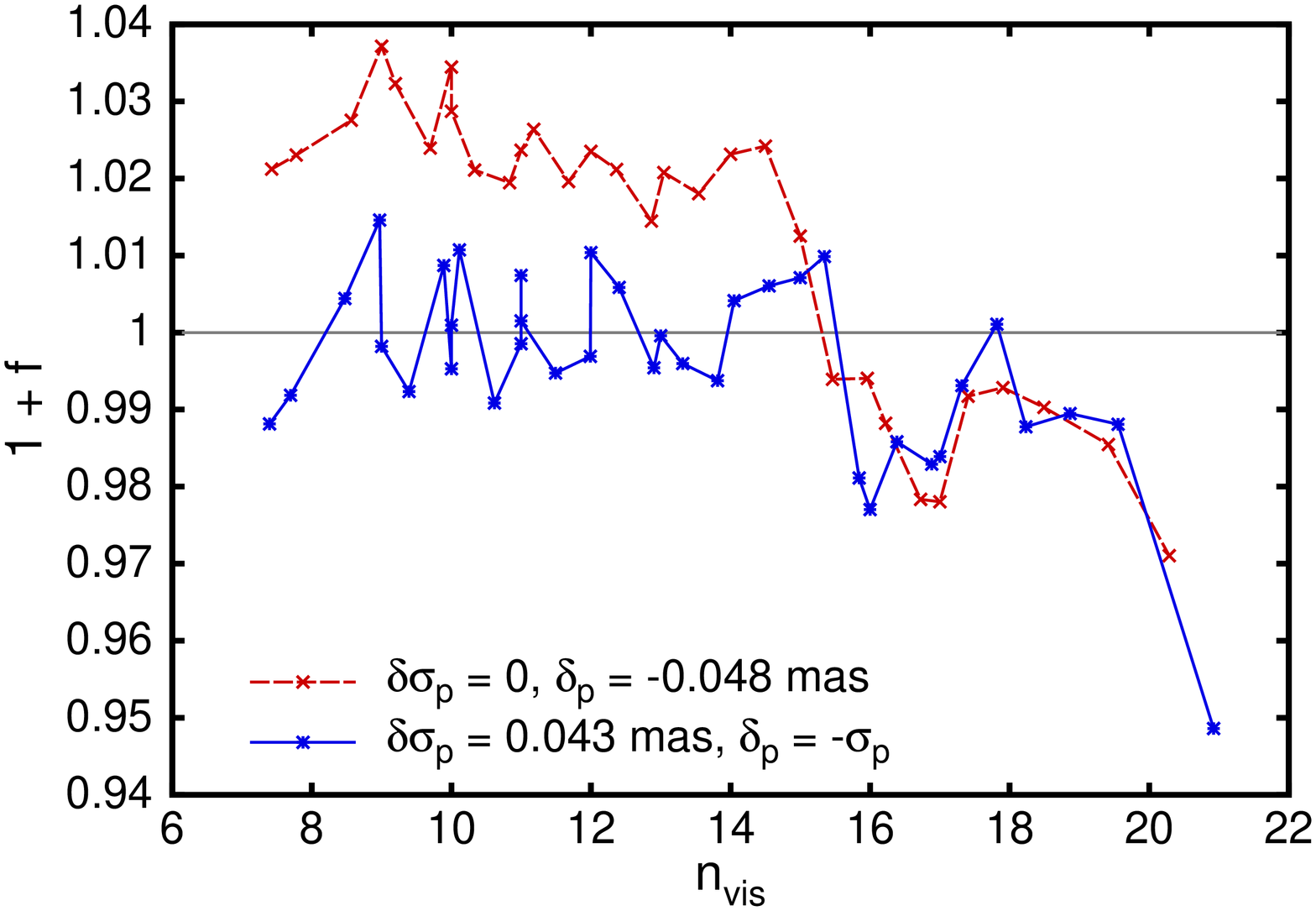,angle=-90,width=\hsize}
\epsfig{file=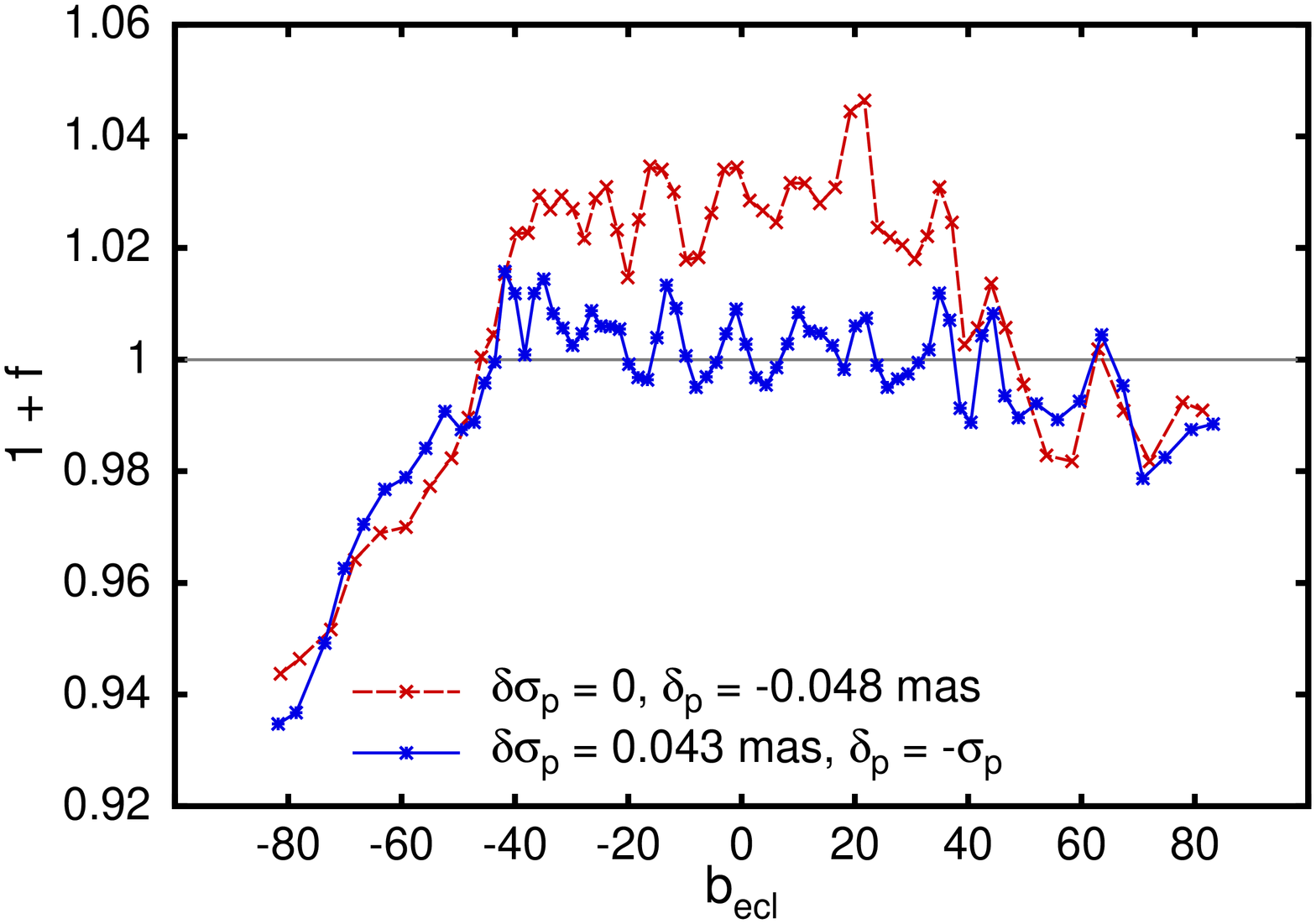,angle=-90,width=\hsize}
\caption{Examining the dependence of the residual distance bias on measurement geometry. In the top panel, we order the sample by the number of visibility periods, $\nvis$, using a sliding mask of width $200000$ moved in steps of $100000$, finding a significant downtrend when using a constant parallax offset (red line). This trend disappears (blue line) when we assume $-\delpar = \sigpar$ and limit $\sigpar < 0.12 \mas$. The last data point contains only $20000$ stars and is likely an outlier. Similarly, the bottom panel shows the systematic distance bias against ecliptic latitude, $\becl$, using a sliding mask of width $100000$, moved in steps of $50000$. The general improvement is as in the top panel. The ecliptic poles imply pencil beams and are thus not reliably measured.}\label{fig:eclatscan}
\end{figure}

\begin{figure}
\epsfig{file=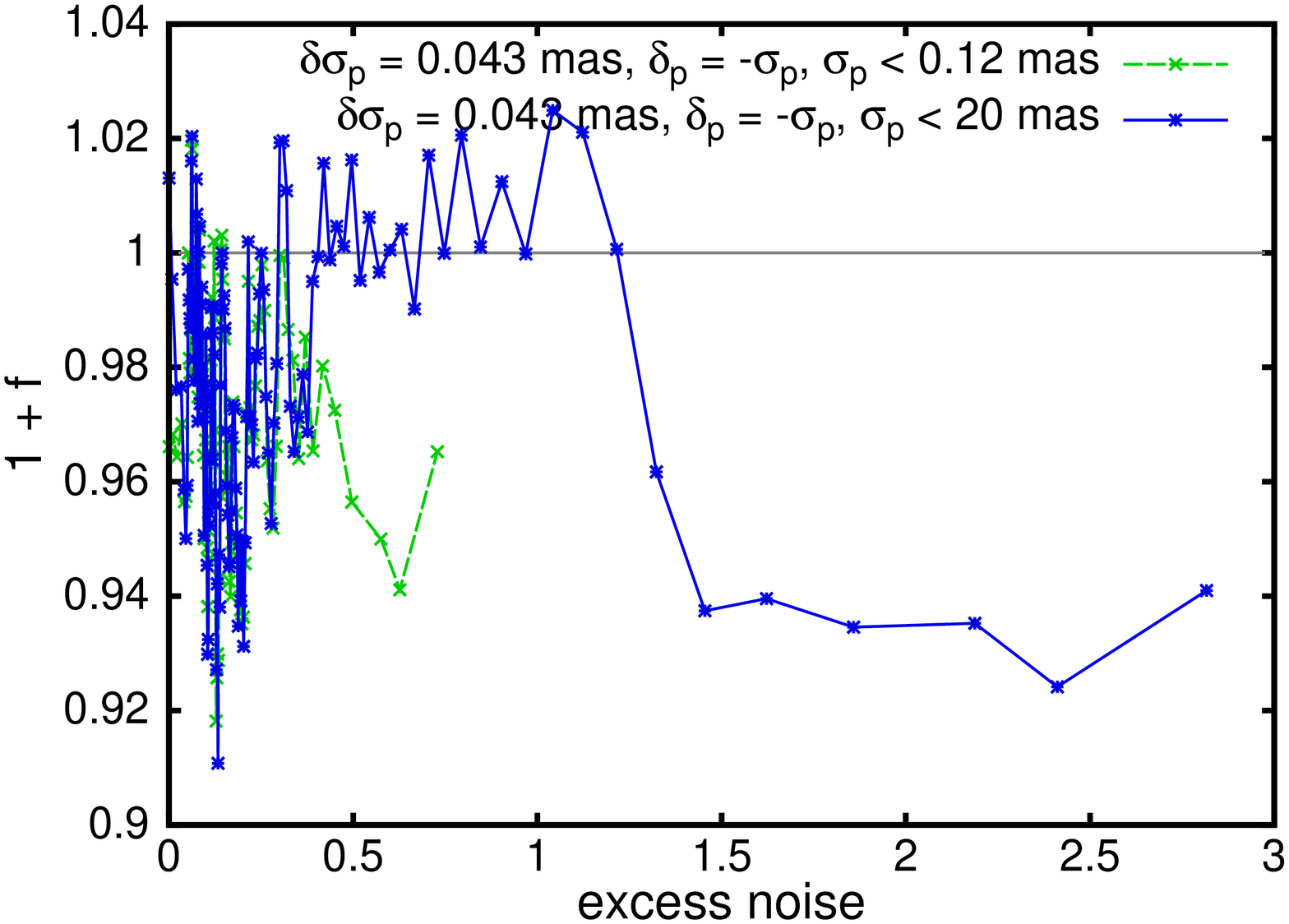,angle=-90,width=\hsize}
\epsfig{file=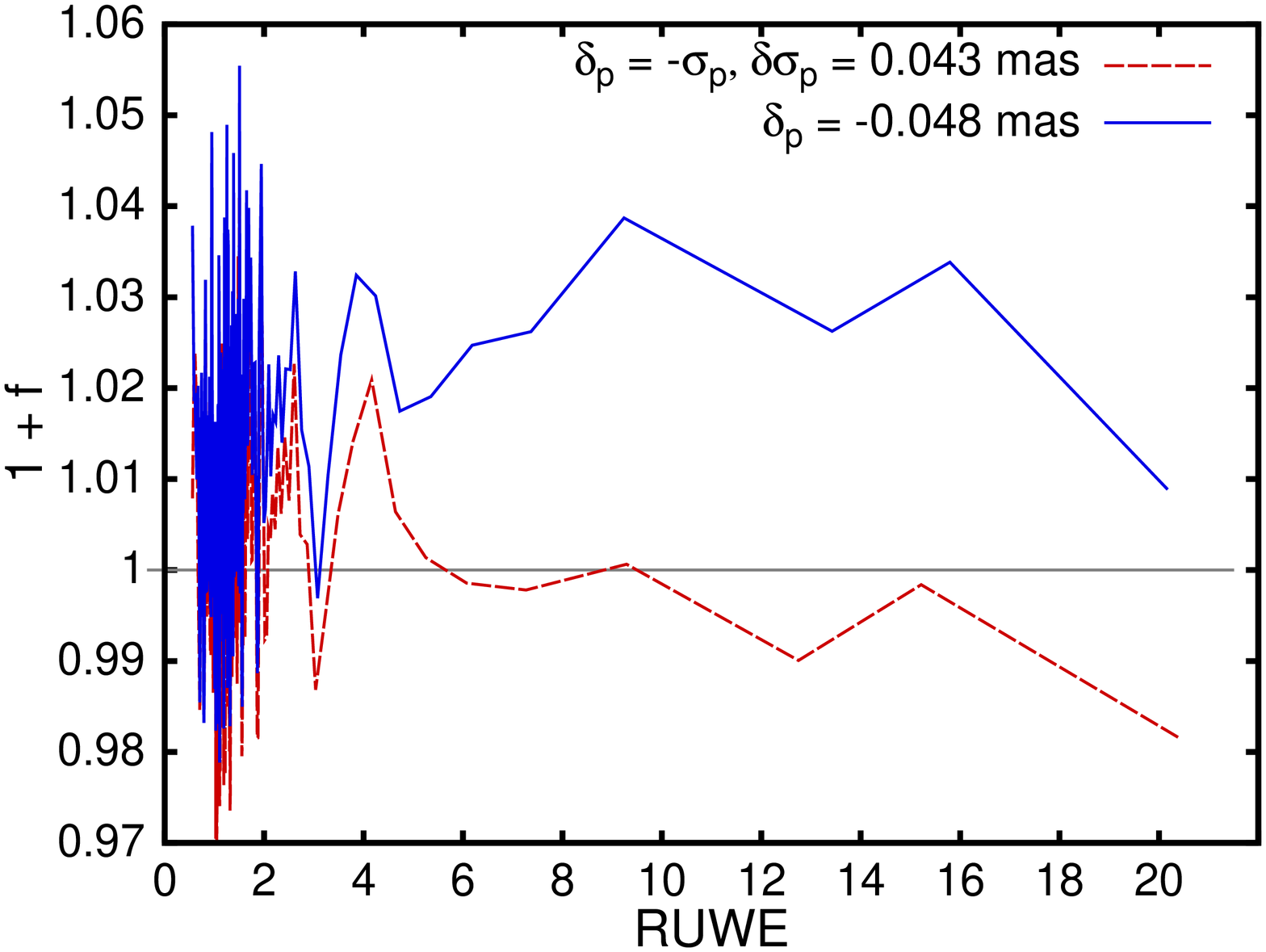,angle=-90,width=\hsize}
\caption{Top panel: Distance bias vs. excess noise. Few stars have a positive excess noise, so we scan in sub-samples of $9000$ stars, moving the mask on the ordered sample by $3000$. Excess noise correlates strongly with the Gaia pipeline's parallax error, so  in this plot we use the assumption $\delpar = \sigma_p$ with $\delta \sigma_p = 0.043 \mas$. The blue line relaxes the cut on parallax quality, which in turn allows us to show the tail of large excess noise values. Bottom panel: distance bias vs. RUWE. No significant trend can be detected when we use the $\delpar = \sigma_p$ assumption, a mild trend exists when $\delpar = -0.048 \mas$, since the RUWE is correlated strongly with $\sigma_p$ and thus high values of RUWE imply a large $\sigma_p$.}\label{fig:excessnoise}
\end{figure}

\begin{figure}
\epsfig{file=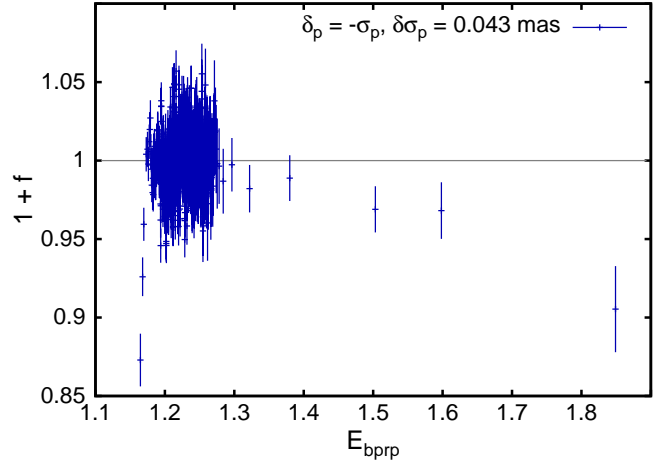,angle=-90,width=\hsize}
\caption{Distance bias $1+f$ vs. the bp-rp excess noise factor. We use the assumption  $\delpar = \sigma_p$ with $\delta \sigma_p = 0.043 \mas$ and scan the ordered sample with a mask of width $12000$ in steps of $4000$ stars each.}\label{fig:bprp}
\end{figure}

\subsection{Bias vs. astrometric parameters}\label{sec:biasastrop}

After discussing the main problems with the sample, it is worth looking at the sample quality vs. the different astrometric parameters. Fig.~\ref{fig:eclatscan} shows the distance bias against the number of visibility periods $\nvis$ (top panel) and the ecliptic latitude $\eclat$ (bottom panel). In both the top and bottom panel we compare the results obtained with the simpler assumption of a constant parallax offset $\delpar = -0.048 \mas$ (red lines) vs. setting $\delpar = -\sigpar$ and limiting $\sigpar < 0.12 \mas$ (blue lines). The $\sigpar$ limit does not significantly alter the results. We were first rather astounded by the strong dependence of the distance bias on $\nvis$ in the naive evaluation (red lines). However, most of this effect is readily explained by the dependence of $\delpar$ on $\sigpar$. First, we note that due to the scanning law of Gaia, $\nvis$ very strongly correlates with $\eclat$, so both trends have a common explanation. The culprit is easily found, when we correct for the apparent dependence of $\delpar$ on $\sigpar$ (blue lines), which diminishes the trend. Consistently with the top panel, most of the dependence of $1+f$ on $\eclat$ in the bottom panel disappears when we apply this correction, with a mild failure of order $f \sim 5 \%$ remaining around the ecliptic south pole. This is far less concerning than it looks since the ecliptic poles imply a pencil beam, where our method loses most of its advantages, and furthermore, the ecliptic poles are in a near-worst case location, since they are close to the Galactic plane almost exactly in the azimuthal direction: this implies weak statistics, and only the $V-W$ correlation term, which carries some mild signal from the galactic warp, no information from the $U-W$ term.  We thus ran a few checks. The distance method did not reveal anything else than near-perfect distance bias when scanning along Galactic $\gl$ and $\gb$ (which would have argued against a failure here), but in turn, we did not see the apparent dip in the bottom panel of Fig.~\ref{fig:eclatscan} when limiting the sample to the more robust stars with low $\Vg$. We further compared the locus of the lower main sequence in absolute G band magnitude for stars with $\eclat \lesssim -55 \degs$ with equivalent fields in Galactic latitude and longitude, finding no significant difference. To summarize all this paragraph: Assuming the dependence of $-\delpar \sim \sigpar$ makes all problems in $\nvis$ and $\eclat$ go away, with some possibility for a problem at $\eclat \lesssim -55 \degs$, so in sensitive studies, one might want to touch these stars with a long pole.

Fig.~\ref{fig:excessnoise} shows the distance bias $1+f$ vs. the astrometric excess noise (top panel) and vs. the RUWE (a quantity based on the chi squared of the astrometric fit and stellar colour). To create these plots we use the usual limits in colour and magnitude, and waived the excess noise limit in the top panel. We note that only a very minor fraction of our stars have a non-zero excess noise value. These stars were censored from the sample. One could expect that at least some stars with larger excess noise should be binaries. In the discussion of \cite{SA17}, the Gaia DR1 sample had a far larger temporal baseline for the astrometry compared to the $\vlos$ measurements, which raises the expectation of an apparent distance underestimate in the statistical distance estimator, since the $\vlos$ measurements would carry the additional velocity dispersion from the binary. Here, the case here is less clear-cut. However, we can still note: stars with a positive excess noise have a slight tendency to distance underestimates in our method (understandable, since the astrometric effect is not clear-cut and we still have some binary dispersion affecting the estimator). If we admit stars with very large $\sigpar$ (which correlates with the excess noise), we find that the sample shows clear signs of a break-down for excess noise values larger than $\sim 1$. Thus, we recommend applying a cut at this value. Further we note, the excess noise measurement is limited to stars with very good signal to noise in Gaia and thus leads to a concentrated sample in distance; neglecting this effect should cause slight distance over-estimates in our selection, so the true distance bias on the large--excessnoise stars is likely slightly underestimated.

The bottom panel of Fig.~\ref{fig:excessnoise} provides our two suggestions for distance evaluation vs. the RUWE as defined in the additional release notes from Gaia. The usual argument is that stars with a large RUWE value are bad and should not be used. However, as we see from the figure, the only trend of $1+f$ that we can detect is in the evaluation where we hold the parallax offset fixed at $\delpar = -0.048 \mas$. However, large values for RUWE (which is an expression for the quality of the astrometric fit) duly correlate with a larger $\sigpar$ given by the Gaia astrometric pipeline. If we correct for the trend in the Gaia parallax offset (green line), this slight bias vs. RUWE vanishes. In short, we cannot see any reason for applying a quality cut in RUWE - of course, stars with large RUWE have worse measurements, but the Gaia pipeline appears to be perfectly fine in not producing any bias vs. RUWE and mirrors the larger uncertainties in larger $\sigpar$ values.

Fig.~\ref{fig:bprp} finally shows $1+f$ vs. the BP/RP flux excess factor $E_{\rm bprp}$. The quantity is often cited as an important quality measure for Gaia data. It compares the flux in the BP+RP bands to the total flux in the Gaia G-band and relies on their very similar coverage. Due to the Gaia passband definitions, $E_{\rm bprp}$ increases on average for red stars, but primarily expresses contamination by neighbouring stars and background, or misidentifications. We note that here we only measure the average distance error, i.e. we are not concerned with single outliers. The result is quite clear-cut: both very low values of $E_{\rm bprp} < 1.172$ and large values of $E_{\rm bprp} > 1.3$ signal compromised stars. We also note, however, that we needed to scan the sample with a very fine sample size mask of $12000$ stars moved in steps of $4000$ stars each, since the number of compromised stars is so low in the sample with $\gb > 10 \degs$. The error distribution within the bulk of the sample shows the usual number of outliers. Scanning with a larger sample size (100000) on the centre of the distribution shows that all sub-samples there have $|f| < 2\%$, with a slightly suspicious region almost exactly at $E_{\rm bprp} \sim 1.2$.

\begin{figure}
\epsfig{file=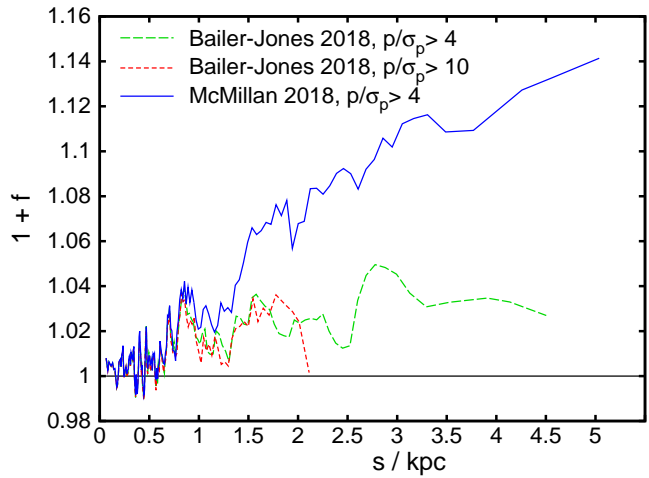,angle=-90,width=\hsize}
\caption{Distance bias $1+f$ vs. distance for the distance sets of Bailer-Jones et al. (2018) and McMillan et al. (2018). We use a sample size of $60000$ stars each, stepping by $20000$.}\label{fig:samplecomp}
\end{figure}

\section{Comparison to previous distance derivations}

It is appropriate to ask how our distances differ from the two previous distance derivations by \cite{BailerJones18} and \cite{McMillan18}, so Fig. \ref{fig:samplecomp} shows a distance scan of $1+f$ vs. their distance $s$. In both cases the distances were estimated under the assumption that the Gaia DR2 parallax zero-point is $-0.029\mas$, based on the quasar results from \cite{Lindegren18}, which was the best available estimate at the times they were computed. We have shown that this is an underestimate for these stars. It is therefore no surprise that Fig.~\ref{fig:samplecomp} shows that both studies overestimated stellar distances substantially, with the typical increase of $1+f$ with distance. In short, due to the parallax offset these distances are compromised and should not be used. The reason why the bias in the \cite{BailerJones18} distances is somewhat smaller than in the \cite{McMillan18} distances is predominantly caused by the fact that we do not have proper expectation values for the \cite{BailerJones18}. Instead, we have only the mode (maximum) of their posterior distance probability distribution, which, due to the skew probability distribution, is significantly smaller, for usual distance PDFs, than the expectation value. While this compensates for some of their intrinsic distance overestimate, the mode infers a difficult-to-predict and variable bias relative to the expectation value of a distribution; also one should not rely on two opposite biases to partly cancel. A longer discussion of this issue can be found in \cite{SA17}. 

\section{The distance to the Pleiades}

Since there has been so much discussion about the distance to the Pleiades, let us quickly analyse the effect that the offset has on their distance estimate. Historically, there {has for a long time been a tension} between a low astrometric distance estimate by Hipparcos \citep[][]{Leeuwen09}, placing the distance of the Pleiades at $s = 120 \pc$ with a formal uncertainty of $1.5 \%$, and results from isochrone fits to their photometry \citep[][]{Meynet93, Stello01}, as well as eclipsing binaries \citep[][]{Zwahlen04, Southworth05}, which placed their distance in the range $\sim (130 - 137) \pc$ with similarly small uncertainties. The Gaia DR2 release \citep[][]{Babusiaux18} estimates the Pleiades' distance at $135.8 \pm 0.1 \pc$. Applying our parallax offset to this estimate brings the distance moderately down to $134.8 \pm 0.2 \pc$, and back towards the average of the stellar-physics-based estimates.

\section{Summary of suggested quality cuts}\label{sec:qualcuts}

Since the previous discussion was rather lengthy with many details, we provide here our suggestions for quality cuts on the Gaia sample, to ensure a minimization of the detected kinematic biases
\begin{itemize}
\item A colour cut $\brp - \rrp < 1.5 \magn$. To be entirely safe, we suggest $0.5 < \brp - \rrp / \magn < 1.4$. 
\item A magnitude cut for $\grp < 14.5 \magn$, and $\brp, \grp, \rrp > 0$. A safer limit is $\grp < 12.5 \magn$ and $\rrp < 13.7 \magn$.
\item $\vpar / \sigpar > 4$, safer is $\vpar / \sigpar > 10$.
\item $\sigpar < 0.1 \mas$ with $\sigpar$ as given by the Gaia pipeline, safer is $\sigpar < 0.07 \mas$.
\item $\nvis > 5$ as pointed out in \cite{Lindegren18} and excess noise $ < 1$.
\item For the BP-RP excess flux factor, use $1.172 < E_{\rm bprp} < 1.3$. Tighter cuts might apply if the number of outliers is important.
\item $s > 80 \pc$ for studies that need assurance of distance systematics $< 4 \%$.
\end{itemize}

Some notes: The $\sigpar$ dependence of the parallax offset is comparably well controlled as long as we choose a safe limit on $\sigpar$. If one needs to use a looser limit on $\sigpar$, we strongly advise to control for the dependence of $\delpar$ on $\sigpar$. If needed, the upper limit on the excess flux factor $E_{\rm bprp}$ can be relaxed a bit with proper caution, see Fig.~\ref{fig:bprp}. As detailed in Section~\ref{sec:biasastrop}, we find no reason to test or cut in RUWE, when applying our set of quality cuts on this sample. Note further that the faint magnitude limit, and likely the colour limit are an imprint of the $\vlos$ measurement quality in the Gaia RV subset, i.e. these should be validated separately for different samples.

\section{Conclusions}\label{sec:Conclusions}

We have used the methods of \cite{SBA} and \cite{SA17} to derive bias-free expectation values for stellar distances in the RV subset of Gaia DR2.\footnote{Please find the datasets with the derived distances and simple estimates of stellar velocities and positions in Galactic coordinates either in the MNRAS online materials or at \url{https://zenodo.org/record/2557803}.}. While Gaia parallaxes have been extensively tested in \cite{Lindegren18}, their quasar sample has no overlap with our study in magnitude, thus the quasar sample is inadequate for testing distances in the important Gaia RV subset for two reasons: quasars have all $\vpar = 0$, i.e. differences in the astrometric pipeline for $\vpar > 0$ cannot be tested, and basically none of their quasars shares the same evaluation cohort as the stellar sample.

We derived Bayesian distances for all stars in the subset and validated the distance expectation values for stars with $\vpar/\sigpar > 4$, the minimum justifiable quality requirement for relative parallax error. All distances and derived kinematics will be made available with this work.

Our study provides clear proof for an average parallax offset $\delpar = -0.054 \mas$ (Gaia parallaxes are too small) with negligible formal uncertainty and a systematic uncertainty of $\sim 0.006 \mas$. The parallax offset is clearly identifiable as such as it results in an almost perfectly linear uptrend of distance bias with distance $s$ that reaches values in excess of $f > 30\%$ for $s > 3 \kpc$.
This offset is comparable to the findings of \cite{Zinn18} using asteroseismic data. It is significantly larger than the value of $-\delpar = 0.029 \mas$ found by \cite{Lindegren18} for quasars. This also advocates a re-analysis of cluster distances. Even the very nearby Pleiades are pulled back by this offset by $\sim 1 \pc$. Similarly, even a comparably benign bias of $10 \%$ creates larger deviations of mean velocities, than found e.g. for the warp/wave pattern in the local Galactic disc. Every study using Gaia DR2 parallaxes/distances should investigate the sensitivity of their results on the parallax biases described here and - for fainter samples - in the DR2 astrometry paper.

We evaluated different assumptions for the parallax error in the Gaia pipeline and found that our estimate for $\delpar$ is nearly unaffected by changing the parallax error. Not adding the additional error $\delta \sigpar = 0.043$ to the astrometric pipeline value $\sigpar$ in quadrature, decreases our estimate for $-\delpar$ by about $0.006 \mas$.

As we used a self-informed prior for the distance-dependent selection function $S(s)$, the method provides a good approximation for $S(s)$, which we provide in equation (\ref{eq:selfunc}). Assuringly, $S(s)$ displays the behaviour expected from population synthesis and the magnitude limits of the RV sub-sample of Gaia: an almost perfectly exponential decrease for $s < 1 \kpc$ related to the main sequence, a knee at intermediate distances, when the magnitude cut passes the level of the subgiant branch, and a slower decrease of $S(s)$ towards large $s$.

After correction for a constant parallax offset we still found a highly significant correlation of the distance bias $f$ with $\sigpar$. While this could point to a problem with the assumptions in our Bayesian distances, this explanation is unlikely since $S(s)$ can be measured to high confidence from the data. To the contrary, our results suggest that $-\delpar$ is roughly proportional to $\sigpar$ (best-fit value $q = 1.05$) after adding the $0.043 \mas$ additional error to the Gaia parallaxes, and further show that stars with $\sigpar \gtrsim 0.1 \mas$ should be discarded from analysis. It is unreasonable to think that the parallax uncertainty is added to the parallax value in a simple way, and so it is no surprise that the required factor is not constant in parallax: $q$ depends significantly on $p$ and around $p \sim 0.09 \mas$, it is closer to $q = 0.5$.  

Resolving this dependency of $\vpar$ on $\sigpar$ also removes a highly significant trend of our measured distance bias with the number of visibility periods $\nvis$, and consequently with ecliptic latitude $\eclat$. We note that our method would still flag a distance problem at the ecliptic south pole. However, when used on a very narrow area on the sky, we lose most of our statistical corrections and do not trust the evaluation. In fact consistent with this expectation, when limiting the sample to the more robust stars with low azimuthal velocity, this dependency was not confirmed, consistent with an evaluation of the derived HR diagram.

We further used the method to evaluate safe limits to be imposed both on apparent magnitude and stellar colour, finding that red stars at $\brp-\rrp > 1.5$ are compromised as well as stars with $\grp \gtrsim 14 \magn$ are flagged for distance underestimates. The most likely explanation is a decline in quality of the otherwise very well determined $\vlos$. 

We further tested for astrometric parameters, finding no biases related to $RUWE$ (after removal of the aforementioned problems), and no strong correlation of $f$ with astrometric excess noise values smaller than $1$. At least for getting distance expectation values in the Gaia RV sample, this strongly argues for not using $RUWE$ as a quality indicator. A mild decrease in distance estimates could point to stellar binaries.

A summary of all quality cuts is provided in Section \ref{sec:qualcuts}.

\section*{Acknowledgements}
We thank our referee, U. Bastian, for very thorough and insightful comments to the paper. It is a pleasure to thank Lennart Lindegren, J. Magorrian, J. Binney, F. van Leeuwen, and A. Mora for helpful comments. RS is supported by a Royal Society University Research Fellowship. This work was performed using the Cambridge Service for Data Driven Discovery (CSD3), part of which is operated by the University of Cambridge Research Computing on behalf of the STFC DiRAC HPC Facility (www.dirac.ac.uk). The DiRAC component of CSD3 was funded by BEIS capital funding via STFC capital grants ST/P002307/1 and ST/R002452/1 and STFC operations grant ST/R00689X/1. DiRAC is part of the National e-Infrastructure. This work has made use of data from the European Space Agency (ESA)
mission {\it Gaia} (\url{https://www.cosmos.esa.int/gaia}), processed by
the {\it Gaia} Data Processing and Analysis Consortium (DPAC,
\url{https://www.cosmos.esa.int/web/gaia/dpac/consortium}). Funding
for the DPAC has been provided by national institutions, in particular
the institutions participating in the {\it Gaia} Multilateral Agreement.

\bibliographystyle{mnras}
\bibliography{paper}
\label{lastpage}
\end{document}